\documentclass{aa}  

\usepackage{graphicx}
\usepackage{xcolor}
\usepackage{color}
\usepackage{soul}
\usepackage{txfonts}
\usepackage[draft]{hyperref}
\usepackage[bordercolor=white,backgroundcolor=gray!30,linecolor=black,colorinlistoftodos]{todonotes}

\def \msun{\ifmmode{{\rm\ M}_\odot}\else{${\rm\ M}_\odot$}\fi}

\begin{document}

   \title{Stripped-envelope core-collapse supernova $^{56}$Ni masses:}

   \subtitle{Persistently larger values than supernovae type II}

 \author{N. Meza\inst{1} \and J.P.~Anderson\inst{1} }

   \institute{European Southern Observatory, Alonso de C\'ordova 3107, Casilla 19, Santiago, Chile}

\titlerunning{SE-SN $^{56}$Ni masses}
\authorrunning{Meza \&\ Andersons}

   \date{Received ; accepted }

 
  \abstract
   {The mass of synthesised radioactive material is an important power source for all supernova (SN) types. \cite{and19} recently compiled literature values and obtained $^{56}\mathrm{Ni }$ distributions for different core-collapse supernovae (CC~SNe), showing that the $^{56}\mathrm{Ni }$ distribution of stripped envelope CC~SNe (SE-SNe: types IIb, Ib, and Ic) is highly incompatible with that of hydrogen rich type II SNe (SNe~II).
   This motivates questions on differences in progenitors, explosion mechanisms, and $^{56}$Ni estimation methods.}
   {Here, we re-estimate the nucleosynthetic yields of $^{56}\mathrm{Ni }$ for a well-observed and well-defined sample of SE-SNe in a uniform manner. This allows us to investigate whether the observed SN~II--SE~SN $^{56}\mathrm{Ni }$
   separation is due to real differences between these SN types, or because of systematic errors in the estimation methods.}
   {We compiled a sample of well observed SE-SNe and measured $^{56}\mathrm{Ni }$ masses through three different methods proposed in the literature. First, the classic `Arnett rule', second the more recent prescription of \citeauthor{kha18}, and third using the tail luminostiy to provide lower limit $^{56}\mathrm{Ni }$ masses. These SE-SN distributions were then compared to those compiled by \citeauthor{and19}.}
   {Arnett's rule - as previously shown - gives $^{56}\mathrm{Ni }$ masses for SE-SNe that are considerably higher than SNe~II. While for the distributions calculated using both the \citeauthor{kha18} prescription and Tail $^{56}\mathrm{Ni }$ masses are offset to lower values than `Arnett values', their $^{56}\mathrm{Ni }$ distributions are still statistically higher than that of SNe~II. Our results are strongly driven by a lack of SE-SN with low $^{56}\mathrm{Ni }$ masses (that are in addition strictly lower limits).
   The lowest SE-SN $^{56}$Ni mass in our sample is of 0.015\msun, below which are more than 25\%\ of SNe~II.}
   {We conclude that there exists real, intrinsic differences in the mass of synthesised radioactive material between SNe~II and SE-SNe (types IIb, Ib, and Ic). Any proposed current or future CC~SN progenitor scenario and explosion mechanism must be able to explain why and how such differences arise, or outline a yet to be fully explored bias in current SN samples.}


   \maketitle
%

\section{Introduction}
One of the big unresolved questions in SN science is the connection of  different types of stellar explosions with possible stellar progenitors. In the case of CC~SNe, that arise from the collapse of the iron core of massive stars, a way to probe the progenitor core structure and the explosion mechanism is to study the nucleosynthesis yields of the explosion. In particular the radioactive yield of $^{56}$Ni is a useful probe of the explosion physics (e.g. \citealt{suw19} and references therein) and it is a fundamental parameter driving the evolution of SN light emission in the first few hundred days (e.g. \citealt{col69}).\\
\indent CC~SNe can be broadly separated into those that show long, persistent hydrogen features in their spectra (hydrogen-rich SNe~II) and those that do not (see \citealt{gal17} for a recent review of SN observational classifications). The latter are broadly classed as `stripped envelope' events (SE-SNe, as above) where the nomenclature refers to the progenitor exploding without most of its outer hydrogen (types IIb and Ib) or additionally helium (Ic) envelopes retained at the epoch of explosion. An analysis into the mass of $^{56}$Ni sythesised in the explosion of SE-SNe is the focus of this paper.\\
\indent When a successful CC~SN explosion occurs, a shock wave propagates outwards and produces explosive nucleosynthesis in the inner parts of the ejecta (composed mainly of silicon and oxygen) that for a short period of time acquires high temperatures of $\sim 5 \cdot 10^9$ K (required to produce the radioactive yields of $^{56}$Ni)\footnote{A recent review on CC~SN explosions can be found in \cite{mul16}. For works on explosive nucleosynthesis see e.g. \cite{arn70,woo73}, together with modern contributions such as \cite{chi17}.}. The shock soon sweeps through the entire ejecta, homologous expansion is achieved and the highly ionized ejecta cools in very close to adiabatic conditions. The early light curve for a hydrogen deficient SE-SN consists of a very fast decline-cooling stage where the thermal emission of the ejecta is highly reduced by the temperature decrease due to expansion cooling. At the same time the ionization of the ejecta is reduced and recombination settles in which may give rise to a short plateau of a few days (e.g. \citealt{ens88,des11}). After this stage the input from radioactive decay, in the form of gamma rays and positrons, becomes fundamental as the heat wave it produces in the inner layers finally encounters the receding photosphere of recombining material, which gives rise to stabilization or increase in the photospheric temperature and the light curve starts to rise to peak. The following light curve behaviour depends on the amount and spatial distribution of the heating material with respect to the bulk of the ejecta. In general the light curve will rise to a peak luminosity that increases with the amount of heating material and will decline in a time scale that depends on the photon diffusion time scale, which grows  with the ejecta opacity and mass. Additionally, the amount of mixing of the heating source also effects the timescales in that for larger mixing (i.e. the radioactive material is found out to a larger radius) the heat wave reaches the outer envelope faster and the gamma ray escape probability will be higher for the ejecta. After peak luminosity, the light curve slope will smoothly approach the slope of the radioactive heating decay, modulo the decreasing deposition function of gamma rays, as it approaches the optically thin regime \citep{des16}. This simple scenario and physical assumptions is the basis for the measurement of $^{56}$Ni masses from observations.\\
\indent Recently, \cite{and19} compiled a large sample of literature $^{56}$Ni measurements for CC~SNe and found significant differences between the $^{56}$Ni distributions of SNe~II and SE-SNe estimated by the community. \citeauthor{and19} argued that such large differences in $^{56}$Ni masses were inconsistent with currently favoured progenitor and explosion models of CC~SNe where significant differences are not expected between the different types.
That work compiled many different estimations from many different authors in the literature and therefore combined a large number of different methodologies. Here, our aim is to investigate whether these recent results are due to errors in the $^{56}$Ni estimation methods, or whether true intrinsic $^{56}$Ni differences do exist between SNe~II and SE-SNe. We do this by concentrating our efforts on $^{56}$Ni estimation in SE-SNe, through compiling a well-defined and well-observed sample of literature SE-SN photometry, and re-estimating $^{56}$Ni masses.
These masses are estimated using different formalisms from the literature and are then once again compared to the $^{56}$Ni distribution of SNe~II. We find that the $^{56}$Ni mass difference persists between SE-SNe and SNe~II.\\
\indent The paper is organised as follows. In the next section we describe our data sample selection. Then in Section 3 we outline how we produce our pseudo-bolometric light curves and the methods used to extract $^{56}$Ni masses from them. Section 4 presents our new SE-SN $^{56}$Ni mass distributions and compares them to SNe~II, then the implications of these findings are further discussed in Section 5. Finally, we present our main conclusions in Section 6.

\section{Data and sample selection} 
\label{sec:data}
To collect the most complete SE-SN sample possible we searched for photometric data listed on the open supernova catalog \citep{gui17}\footnote{https://sne.space/}.
Besides using publications of individual SNe, we used samples from the Harvard-Smithsonian Center for Astrophysics (CfA) \citep{bia14}, the Carnegie Supernova Project (CSP-I) \citep{tad18a}, and the Sloan Digital Sky Survey (SDSS) SN survey II \citep{tad15}. The wavelength coverage and cadence of the photometric observations is highly varied, from just a couple of optical-wavelength observations at a few epochs, to a wide coverage from the NUV to the NIR and with several observations from close to explosion out to nebular phases, more than 100 days post peak luminosity. This full compilation comprises 133 SE-SNe. 
From this initial sample, we selected SE-SNe 
with coverage in the optical $BVRI$ (or SDSS analogues $gri$) and near-infrared (NIR) $YJH$ bands\footnote{To be included, a SN does not need to have photometry in all of these optical and NIR bands, but that the photometry extends from $B$ on the blue side and to $H$ on the red.}.
Then, objects are also removed that do not have photometry covering phases close to the peak. SE-SNe are retained if they have at least one data point before and after the estimated peak luminosity in the bolometric light curve (see below our procedure to obtain the bolometric light curves). 
For an individual SN, if there was more than one source of photometry in a given band, we checked whether they were consistent. If they were significantly different considering the errors, and if it was not possible to determine the reason (e.g. different zero-points or photometric systems) the discrepant source with less data was simply removed.
Following the above selection criteria we obtain a sample of 37 SE-SNe (listed in Table~\ref{tab:table}).
\\
\section{Methods}
\label{sec:analysis}
Here we first outline how we produce bolometric light curves for our sample and then summarise the different methods used to estimate $^{56}$Ni masses.

\subsection{Bolometric light curves}
\begin{figure}[ht!]
 \centering
\includegraphics[width=\linewidth]{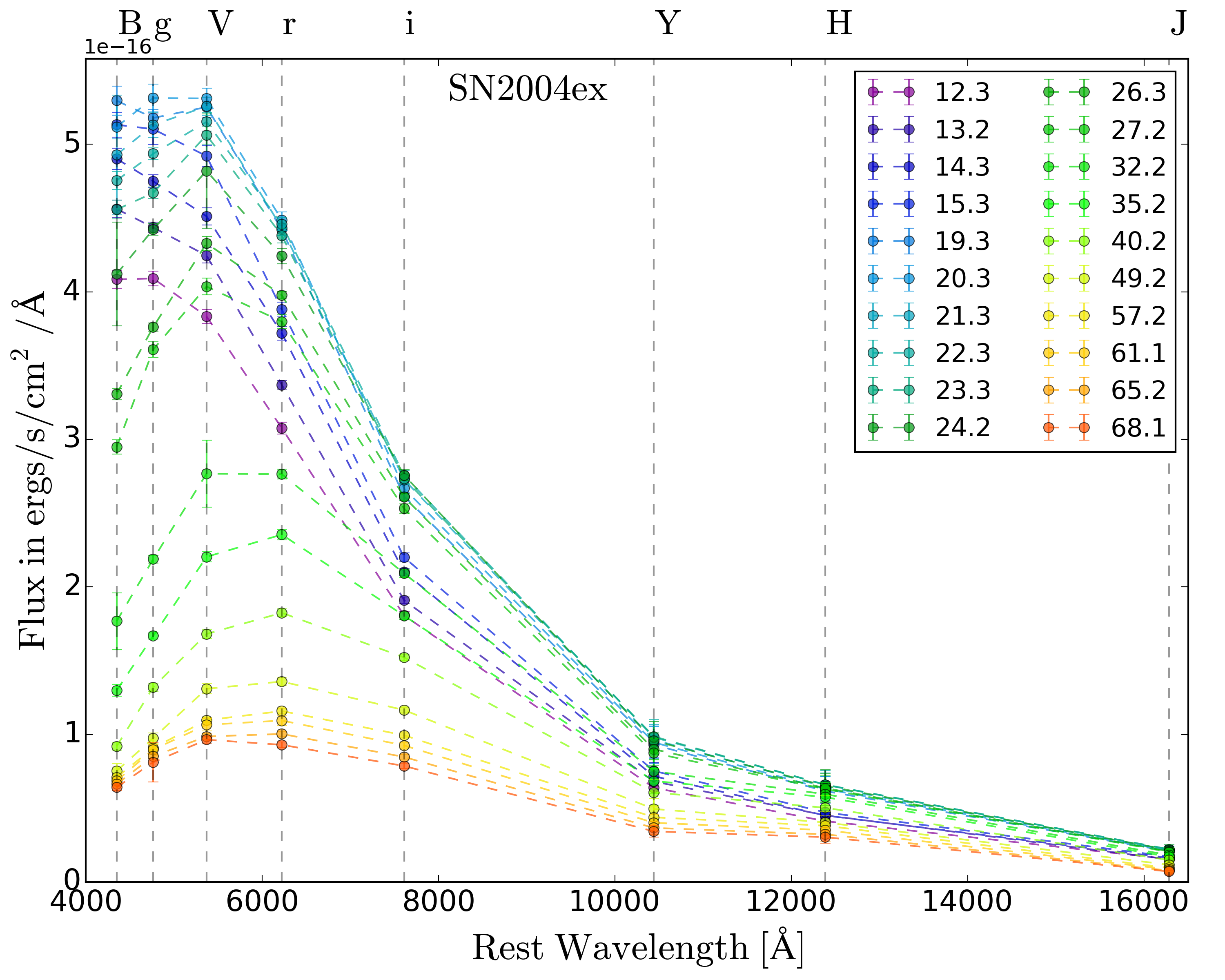}
\caption{Example of the temporal SED evolution for the type IIb SN~2004ex. Each SED is colour coded by the time since explosion (listed in the legend, in days). Each filter is labeled with a dashed vertical line and the name of the filter is given on the top axis.\label{fig:SED_example}}
\end{figure}

 After compiling the photometry, the steps used to obtain bolometric light curves for each SN were as follows: 
\begin{itemize}
    \item We first selected the time window $(t_{min},t_{max})$ within which the SN has photometry in those bands within the $B$ to $H$ range. This conservative interval was chosen so as to avoid data extrapolation. Only in the extreme case that a photometric band barely covered the peak, did we linearly extrapolate the data up to a maximum of 7 days, including an extrapolation error of 0.5 magnitudes. If there was a band that limited too much the final time range of the bolometric light curve, the band was removed: except for the case of $B$ or $H$ photometry where their removal would shorten the wavelength range.
    \item  Given the time window from above, we chose to interpolate all photometry to the epochs of the filter having the most homogeneous coverage. To define this filter, we used the entropy measure of the coverage distribution (i.e. the histogram/distribution of the epochs observed, ${p_k}$) of each filter, which is calculated by the Shannon formula, 
    \begin{equation}
        S = -\sum_k p_k \cdot log(p_k)
    \end{equation}
    and the filter with the maximum entropy value was chosen. We then linearly interpolated all the other bands to the epochs of the filter as defined above, providing us with magnitudes evaluated in the same baseline $m(\{t_i\})$.
    \item Before converting to flux we corrected the magnitudes for the Galactic reddening \citep{sch11} using a standard Cardelli extinction law with $R_V = 3.1$. Although probably uncertain (see later discussion), we also corrected for host galaxy extinction using values from the relevant references (see Table~\ref{tab:table}).
    \item All photometry was transformed to the AB system and flux densities were calculated using standard formulae. An example resulting spectral energy distribution (SED) is shown in Figure \ref{fig:SED_example}.
    \item Fluxes $F_{\nu}(t_i)$ were integrated in frequency space using Simpson's rule. With this integrated flux $F(t_i)$ the luminosity was then obtained using the luminosity distance, 
    \begin{equation}
        L(t_i) = 4\pi d_L^2 F(t_i)
    \end{equation}
    and we do not extrapolate the fluxes outside our defined wavelength range of $B$ to $H$. Therefore, our resulting light curves should be considered pseudo-bolometric, and are a lower limit to the true bolometric luminosity at all times.
\end{itemize}

\begin{figure*}[ht!]
 \centering
\includegraphics[width=\linewidth]{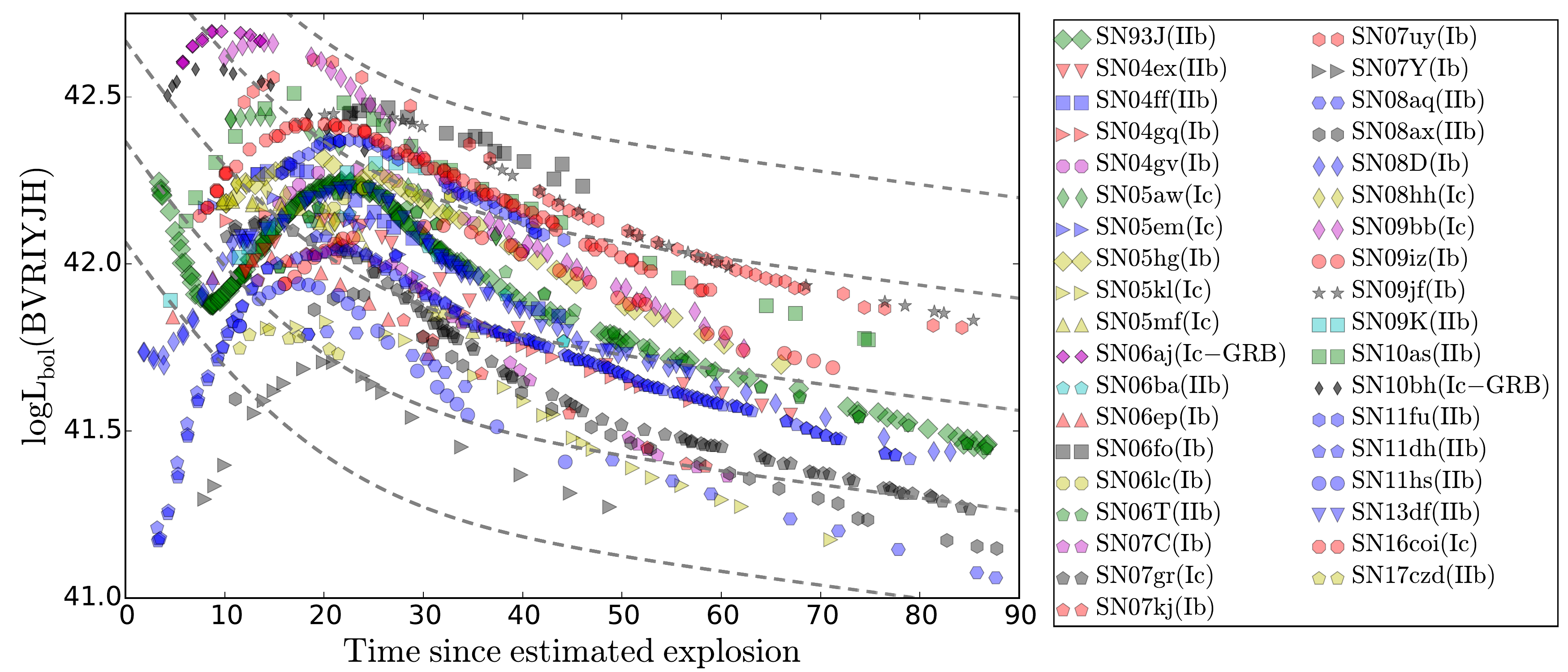}
\caption{$BVRIYJH$ bolometric light curves for our sample of 37 SE-SNe. SN types are listed in the legend. The dashed grey curves show reference $^{56}$Ni decays, with a factor of two in $^{56}$Ni mass separation. From these reference slopes its can be seen that all SE-SNe within our sample decline steeper than the $^{56}$Ni decay after $\sim 60$ days past explosion. \label{fig:allbolos}}
\end{figure*}

Our $BVRIYJH$ bolometric light curves for the full sample are presented in Fig.~\ref{fig:allbolos}. This sample has a total 14 SNe~IIb, 13 SNe~Ib, 6 normal SNe~Ic, 2 SNe~Ic-BL and 2 SNe~Ic-GRB. While the calculated values for these two SNe~Ic-GRB are presented in this paper, they are not included in the comparison to the SN~II population. This is because these objects show the largest degree of asphericities among SE-SN \citep{mae08}, and their emission may also have a significant contribution from sources other than radioactive decay \citep[e.g.][]{Wang2017,Lu2018}.
To obtain peak luminosities we applied a local polynomial regression with a Gaussian kernel, using the public modules from PyQt-fit in Python\footnote{https://pyqt-fit.readthedocs.io/en/latest/index.html}. The smoothing function obtained is sampled in a dense grid and the maximum is obtained directly from this grid. Using explosion epochs as estimated in literature references (see Table \ref{tab:table}), we obtained the rise time ($t_{rise}$ or $t_{peak}$) distributions that are presented and discussed in Appendix~\ref{sec:A-1}. 

\subsection{Methods to measure $^{56}$Ni masses\label{nimassest}}
We now outline the three methods used to estimate $^{56}$Ni masses for our SE-SN sample. To date, the most commonly used method in the literature is the application of the so called ``Arnett rule'' \citep{arn82}. This ``rule'' is derived analytically from a simple model with several assumptions (see \citealt{Khatami19} for a discussion of the assumptions involved). The rule states that the luminosity at peak, $L_{peak}$ is equal to the instantaneous power from radioactive decay $L_{heat}$, which for the case of $^{56}$Ni decay can be written as, 
\begin{equation}
    L_{peak} = L_{heat}(t_{peak}) = \frac{M_{Ni}}{M_\odot}\left[(\epsilon_{Ni}-\epsilon_{Co})e^{-t_{peak}/t_{Ni}}+ \epsilon_{Co}e^{-t_{peak}/t_{Co}}\right] 
    \label{eq:decay}
\end{equation}
where $\epsilon_{Ni}=3.9\cdot 10^{10}$ erg/s/g, $\epsilon_{Co}=6.78\cdot10^9$ erg/s/g, $t_{Ni}=8.8$ days and $t_{Co}=111.3$ days (with $^{56}$Co being the daughter product of $^{56}$Ni). 

The other established method to estimate $^{56}$Ni masses uses the bolometric luminosity at nebular phases, when the ejecta is optically thin. In this phase gamma rays are expected to be fully or partially trapped in the ejecta and the reprocessed energy is released entirely. With the assumption that the gamma rays are fully trapped in the ejecta the bolometric light curve should follow equation \ref{eq:decay} and obtaining the $^{56}$Ni mass is trivial, measuring the bolometric luminosity at a given epoch in the radioactive tail. However, this formalism has generally not been applied to SE-SNe in the literature, while it is the standard method to estimate SN~II $^{56}$Ni masses. In the case of SNe~II, light-curve tail decline rates generally follow that predicted by $^{56}$Co decay (see e.g. \citealt{and14a}). However, this is not the case for SE-SNe (as can be seen in Fig.~\ref{fig:allbolos}). Still, even though SE-SN light curves decline quicker than the predicted rate, the Tail can be used to estimate a lower limit for the $^{56}$Ni mass, and that is what we do here. To ensure that we use luminosities when the light curve has entered the nebular phase, we selected epochs in the bolometric light curve that fulfilled the condition:
\begin{equation}
    t> \min (t_{peak}+2t_{half},t_{peak}+25)
\end{equation}
where $t_{half}$ is the time taken for the light curve to decline from peak luminosity to half that value. The last three of these epochs were then fitted with the relation in equation (\ref{eq:decay}) giving our lower limit $^{56}$Ni mass.

\begin{figure}[ht!]
 \centering
\includegraphics[width=\linewidth]{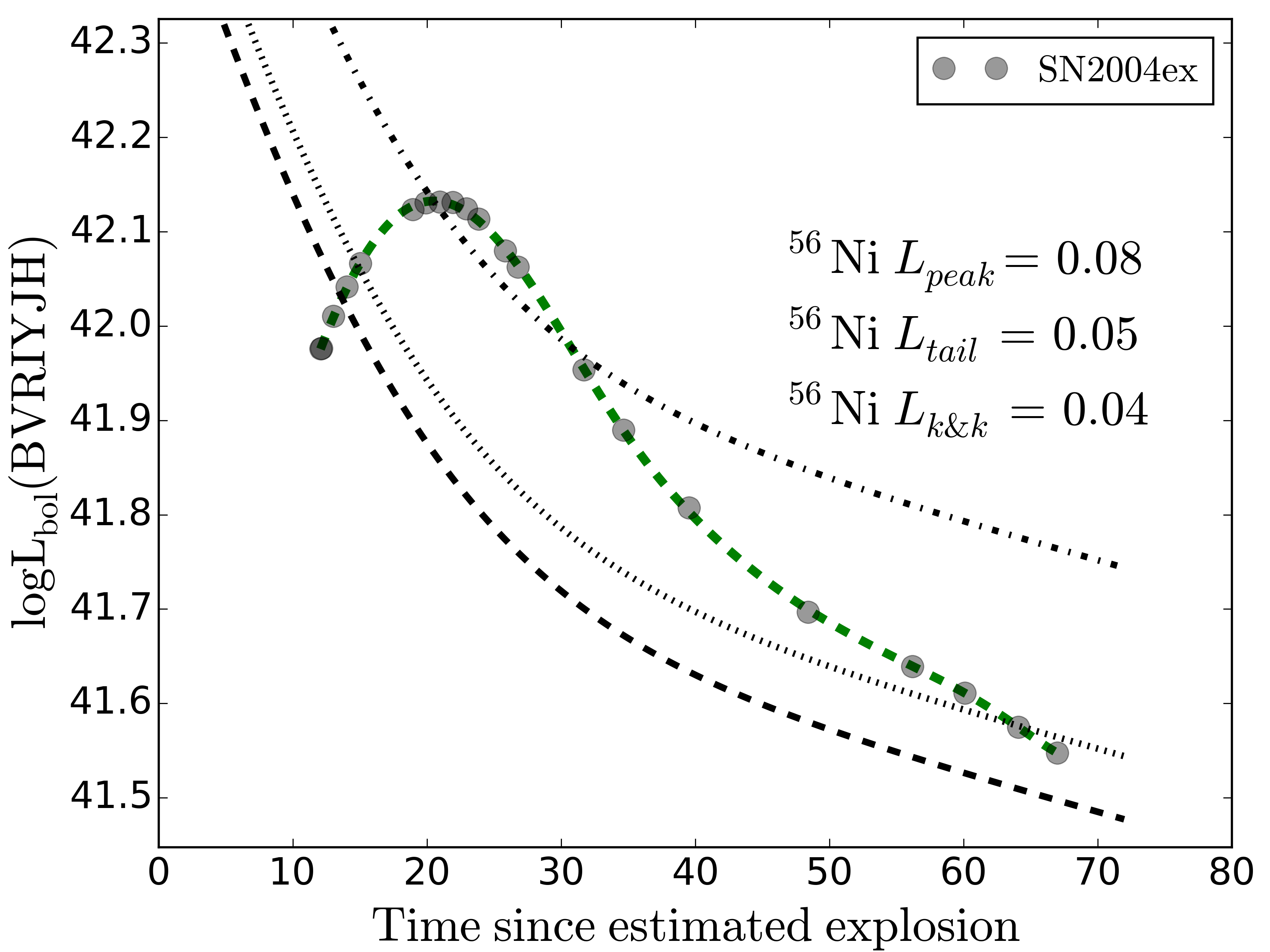}
\caption{Example light curve and analysis results for the well sampled $BVRIYJH$ bolometric light curve of SN~2004ex. The dashed-dotted, dashed, and dotted lines marks the $^{56}$Ni decay, assuming an initial $^{56}$Ni mass given by Arnett's rule, the \cite{Khatami19} method (dubbed ``k\&k''), and the lower limit from the Tail respectively. The fit used to obtain the maximum of the light curve is shown as a green dashed line.\label{fig:bol_example}}
\end{figure}

Recently, a new method was proposed to measure (amongst other things) the $^{56}$Ni masses in SNe \citep{Khatami19}. Here, $^{56}$Ni masses are derived through an integration of the equation for the global energy conservation, using several assumptions on the temporal behaviour of the heating source and internal energy. The \citeauthor{Khatami19} formalism gives rise to the simple relation:
\begin{equation}
    L_{peak} = \frac{2}{\beta^2 t_{peak}^2}\int_0^{\beta t_{peak}}{tL_{heat}(t)dt}
\end{equation}
$\beta$ parameterises the degree of mixing and changing opacity in the ejecta. To calculate $^{56}$Ni masses for our SE-SN sample using this relation we use the suggested values of $\beta$ in \cite{Khatami19}, from their Table 2: 0.82 for SNe~IIb, and $9/8$ for SNe~Ib and SNe~Ic (including SNe~Ic-BL)\footnote{This SNe~Ib/c $\beta$ value is calibrated from models presented in \cite{des15b,des16}. Those models present quite a low degree of mixing.}.
We note here that there does not appear to be strong evidence for the use of these specific $\beta$ values. Indeed, assuming one $\beta$ value for SE-SNe of different ejecta mass may be too simplistic.\\
\indent When possible, $^{56}$Ni masses are derived for all SE-SNe in our sample through the three procedures outlined above. Fig.~\ref{fig:bol_example} shows an example well-sampled bolometric light curve (SN~2004ex) with the $^{56}$Ni decay curves plotted derived from the synthesised  $^{56}$Ni mass from each procedure. 
The distributions resulting from all SE-SNe used in this work are now discussed in detail.

\section{Results}
\label{sec:Results}
\begin{figure}[ht!]
 \centering
\includegraphics[width=\linewidth]{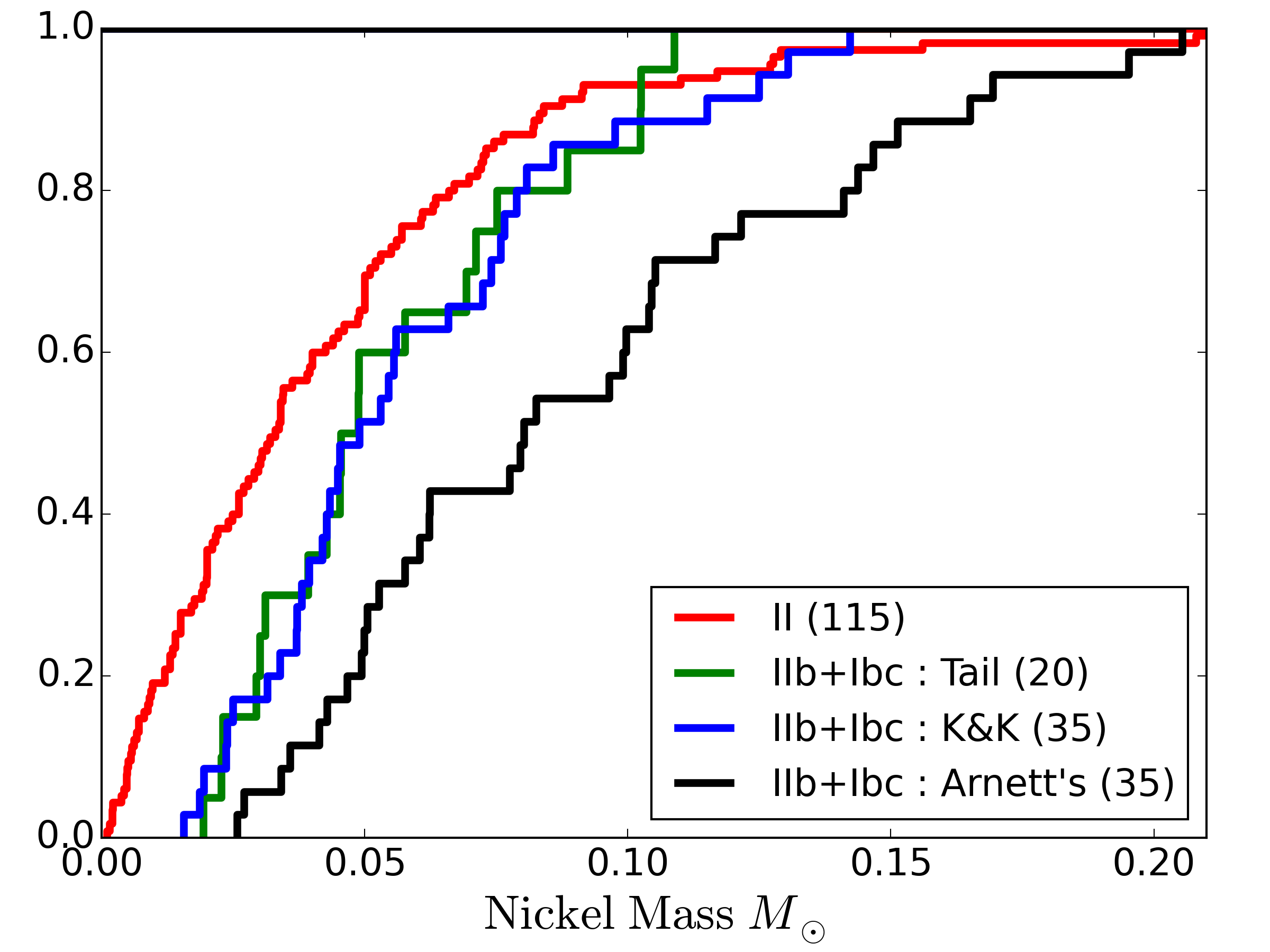}
\caption{Cumulative distributions of SE-SN $^{56}$Ni masses derived through the three methods outlined in the text compared to that of SNe~II \citep{and19}. \label{fig:hist_all}}
\end{figure}

\begin{figure}[ht!]
 \centering
\includegraphics[width=\linewidth]{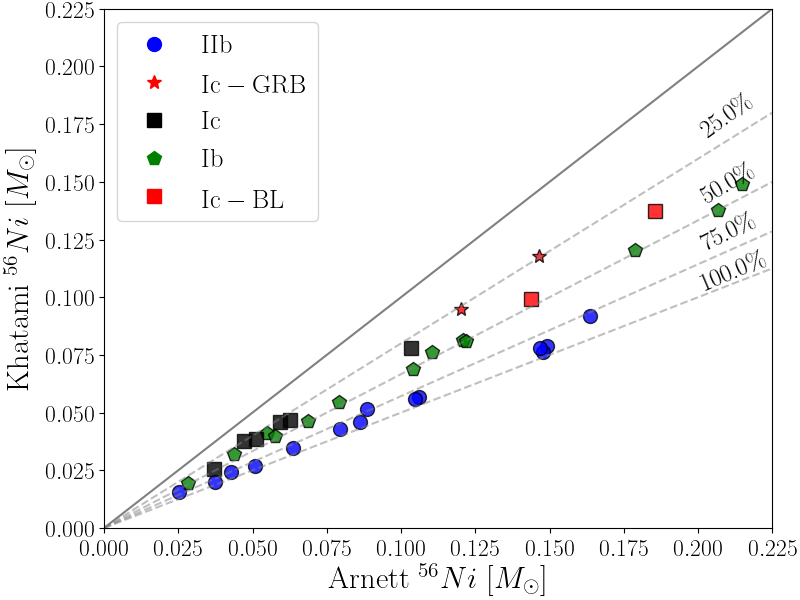}
\caption{Comparison of the $^{56}$Ni masses as measured by Arnett's rule and the \citeauthor{Khatami19} prescription. The solid diagonal line shows the one to one relation between the two methods and the dashed lines show different percentual differences between them. \label{fig:Ni56_comp}}
\end{figure}

Armed with the $^{56}$Ni masses calculated in the previous section, we now compare the SE-SN distributions derived from the different methods and contrast these with the SN~II distribution from \cite{and19}. In Fig.~\ref{fig:hist_all} we show cumulative distributions for the $^{56}$Ni masses analysed in this work, where we combine all SE-SN types (IIb, Ib and Ic - but not Ic-BL). 
It is clear that the $^{56}$Ni masses obtained using the \citeauthor{Khatami19} and Tail methods are significantly lower than those estimated through Arnett's rule. However, Fig.~\ref{fig:hist_all} also shows that the SE-SN $^{56}$Ni mass distributions from the first two methods still appear to be shifted towards higher values than that of the SN~II. It is also clear from Fig.~\ref{fig:hist_all} that there are many SN~II $^{56}$Ni masses below the lowest SE-SN value (through all estimation methods). This point is discussed further in the next section.\\
\indent In Table \ref{tab:KS_table1} the results of a two sample Kolmogorov-Smirnov (KS) and Anderson-Darling (AD) tests between the distribution of $^{56}$Ni masses of SNe~II and SE-SNe are presented. 
The result from \cite{and19} is recovered here - for a smaller sample of SE-SNe, but using Arnett's rule - in that SE-SNe present significantly higher $^{56}$Ni masses than SNe~II. While for the Tail and \citeauthor{Khatami19} methods the significance of this difference is reduced, the difference still persists: SE~SNe in our sample\footnote{Which we have no reason to believe is unrepresentative of the \textit{observed} sample of SE-SNe in the literature.} produce more $^{56}$Ni in their explosions than SNe~II. We reiterate here: the Tail $^{56}$Ni masses are lower limits given that for the majority of SE-SNe their tail luminosities decline more quickly than the rate predicted by $^{56}$Co decay (Fig.~\ref{fig:allbolos}) implying significant escape of the radioactive emission. We also note that Fig.~\ref{fig:hist_all} suggests that the $^{56}$Ni masses derived from the \citeauthor{Khatami19} method and those from the Tail are more or less the same. Given that the latter are lower limits this implies that the former are probably \textit{underestimated}, suggesting that our employed $\beta$ values are possibly in error (see additional discussion below).
We do not attempt to evaluate differences between different SE-SN sub types here, given the low number of objects in each class.\\
\indent Figure \ref{fig:Ni56_comp} presents a comparison of the $^{56}$Ni masses derived through Arnett's rule and the \citeauthor{Khatami19} method. The largest difference between the methods is for the SNe~IIb, in that Arnett's rule gives $^{56}$Ni masses that are around twice as large as those of \citeauthor{Khatami19}. In the case of SN types Ib, Ic and Ic-BL Arnett generally gives values that are 25-50\%\ higher than \citeauthor{Khatami19}. Finally, for our sample of well-observed SE-SNe presented here there are no $^{56}$Ni masses larger than $\sim 0.2$ $M_\odot$ (0.21$M_\odot$ for the type Ib SN2007uy, using Arnett's rule). \cite{and19} discussed the existence of a tail of $^{56}$Ni masses for SE-SNe out to extremely high values nearing 1\msun. The lack of such high values in the current SE-SN sample (that is well sampled in both wavelength and time) suggests that maybe those extreme literature values are in error due to a lack of observational data and/or errors in corrections such as those for host galaxy extinction. Alternatively, it is possible that those SE-SNe with high $^{56}$Ni masses arise from distinct explosion mechanisms where radioactive decay is not the dominant luminosity source (see later discussion).\\
\indent Before discussing the implications of our findings, in the next subsections we discuss possible systematics in $^{56}$Ni mass estimations and how these may affect our results.\\

\begin{table*}
\centering
\caption{Two sample KS and AD (two sided) tests between the distribution of $^{56}$Ni masses of SNII (115 SNe) and SE-SNe (35 for Arnett, 35 for \citeauthor{Khatami19}, and 20 using the Tail).
\label{tab:KS_table1}}
\begin{tabular}{c|cc|cc}
\hline
&  Kolmogorov-Smirnov test & & Anderson-Darling test &\\
\hline
{Method} & {p-value}  & {D} & {p-value}  & {A}\\
\hline
Tail 
& 0.039   & 0.334   & 0.031   & 2.529 \\
\citeauthor{Khatami19}
& 0.002   & 0.351 & 0.001   & 7.197  \\
Arnett
& 6.694E-07  & 0.506 & $<$1.000E-03   & 20.017 \\
\hline 
\end{tabular}\\
\end{table*}

\subsection{Systematics}
While each $^{56}$Ni mass estimation method described in Section \ref{nimassest} has its own caveats, all three are susceptible to uncertainties in (1) the explosion epoch $t_0$, (2) the bolometric correction used to extrapolate the missing flux, and (3) the employed host galaxy reddening values. Here, as outlined above, we do not use any bolometric correction.
However, we only include SE-SNe in our sample that have data between $B$ and $H$ bands\footnote{Regardless, our photometry should cover $\approx 80 \%$ of the flux at the epochs used in this work \citep{lym16}.}, and this selection criteria is applied consistently across the sample. While this removes the uncertainty of calculating bolometric luminosities from only a few optical photometric points (as has been done in previous works), it means that our estimated luminosities are `pseudo-bolometric' and are lower limits to the true luminosity at any epoch. These pseudo-bolometric luminosities therefore translate to lower limits to estimated $^{56}$Ni masses. However, this is not a problem for the main aim of this work. This work aims at testing whether there exist true differences in $^{56}$Ni masses between SE-SNe and SNe~II and in the previous subsection we conclude that indeed true, intrinsic differences persist in that SE-SNe produce more radioactive material than SNe~II. Thus, our decision to not correct for the missing flux outside the $B$ and $H$ bands only reinforces our result: making full bolometric corrections would produce higher SE-SN $^{56}$Ni masses and therefore produce even more statistically significant $^{56}$Ni mass differences than we present.

\subsubsection{Explosion epochs}
\begin{figure}[ht!]
 \centering
\includegraphics[width=\linewidth]{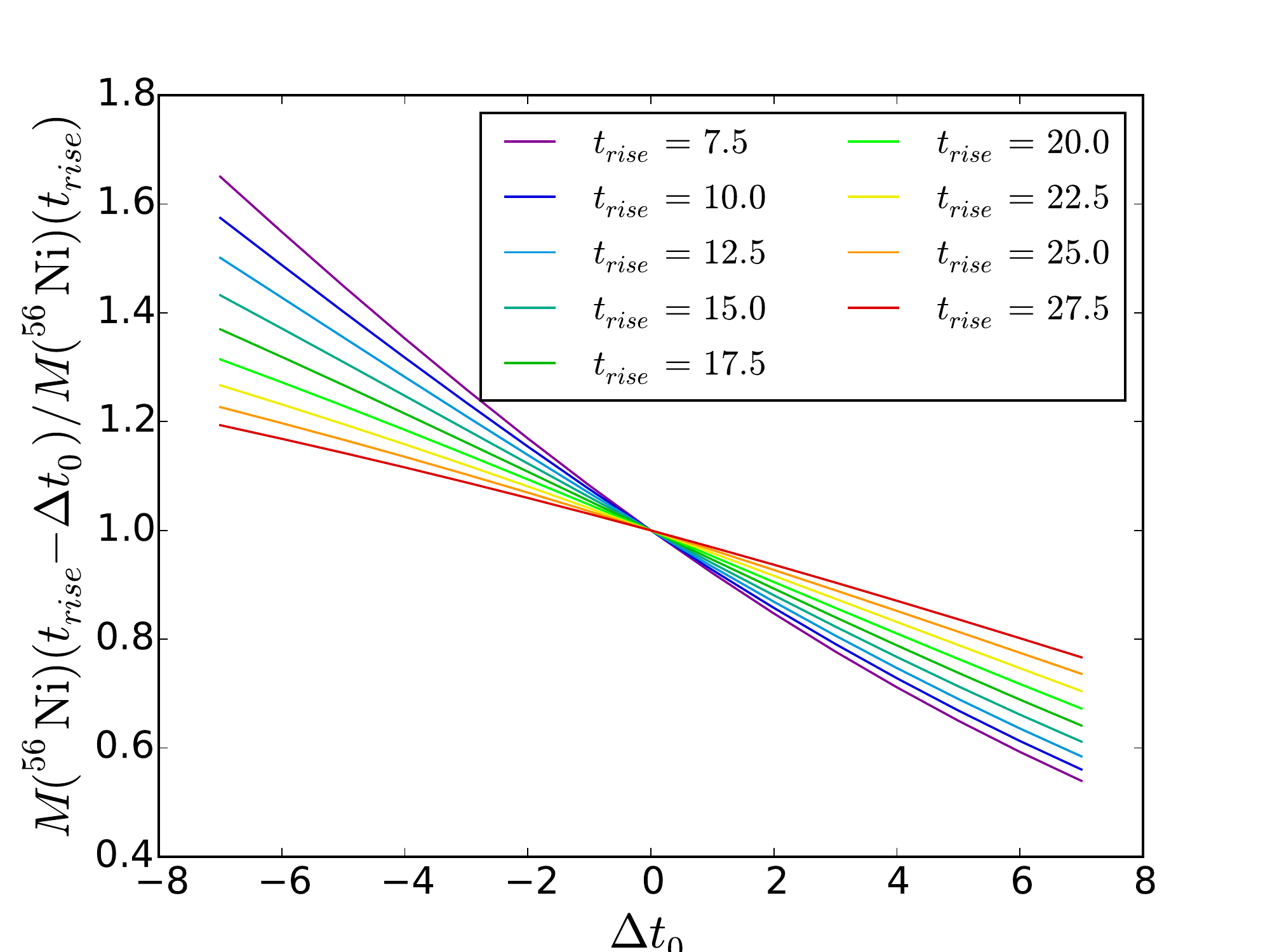}
\caption{Fractional variation of the $^{56}$Ni mass using the Arnett rule, as a function of the rise time variation in days. The shorter the real rise time the more the $^{56}$Ni mass is affected by the unknown explosion epoch.\label{fig:peak_shift}}
\end{figure}

The effect of the uncertainty on the explosion epoch can be more easily tested. In Fig.~\ref{fig:peak_shift} we show the fractional difference in the $^{56}$Ni mass, changing the explosion epoch and using Arnett's rule, for different intrinsic rise times. $^{56}$Ni masses are increased with longer rise times. When changing the rise time $\pm 7$ days, the $^{56}$Ni mass variation goes from $\pm 20\%$ for longer rise times of $\sim 20$ days, typical for a SN~IIb, to $\pm60\%$ for very short rise times of $\sim 10$ days, similar to SNe~Ic.\\
\indent Following the above, we test the dependence of our results on the uncertainty of our employed explosion epochs using the most extreme scenario possible: we redefine explosion epochs to be just one day before the discovery epoch. This effectively reduces the rise time to it's minimal value and therefore pushes the $^{56}$Ni mass to it's minimal value (through this systematic). 
Re-calculating $^{56}$Ni masses using these extreme explosion epochs, and again running a KS test on the SE-SN and SN~II distributions, we find p-values of $\sim$10\%\ for \citeauthor{Khatami19}, while the Tail and Arnett give 5\% and 0.5\%, respectively. Thus, while the statistical significance of differences in $^{56}$Ni mass between the SN types is lessened (as of course expected), the difference still persists.\\
\indent We emphasize that our extreme approach considers the very unlikely possibility that the discovery epochs are within a day of SN explosion. Therefore, we conclude that our results and conclusions are robust to explosion epoch uncertainties.

\subsubsection{$^{56}$Ni mixing}
\begin{figure}[ht!]
 \centering
\includegraphics[width=\linewidth]{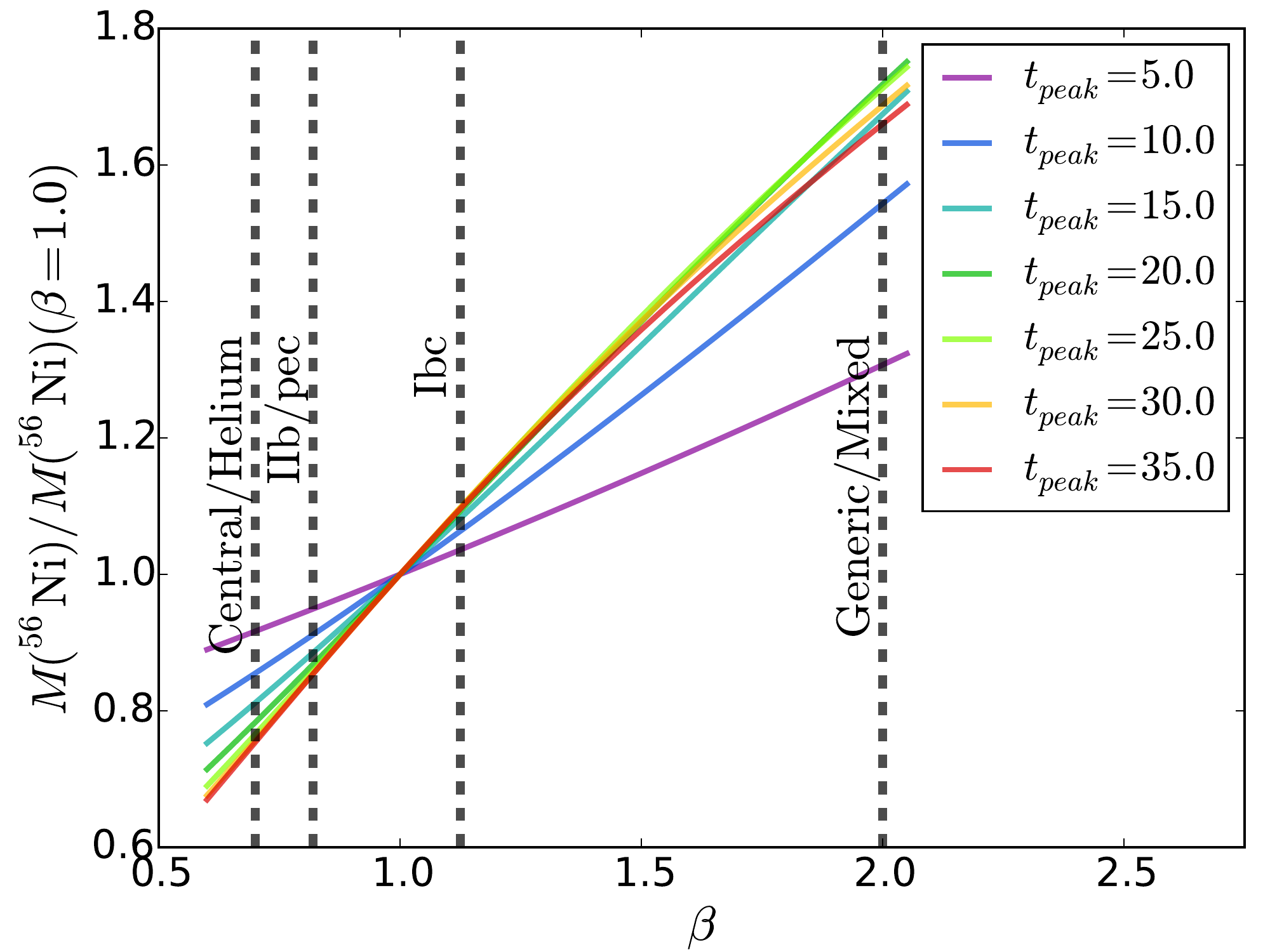}
\caption{Fractional variation of $^{56}$Ni masses using \citeauthor{Khatami19}, as a function of $\beta$. Dashed black lines reference $\beta$ values from different progenitor structures (where labels show which values are suggested for the different SNe). Curves for different initial rise times are displayed in different colours.\label{fig:khatami_shift}}
\end{figure}

The \cite{Khatami19} $^{56}$Ni mass estimation method employs `$\beta$' that paramterises the amount of $^{56}$Ni mixing in the ejecta, which has a strong influence on the resulting $^{56}$Ni mass. 
In Fig.~\ref{fig:khatami_shift} we show the fractional $^{56}$Ni mass variation as a function of $\beta$ for SNe within different rise times. Suggested values for typical progenitor structures and composition are also shown. For rise times greater than 10 days, $^{56}$Ni masses increase by up to a factor two higher when changing $\beta$ from $\sim$0.6 to $\sim$2.0. As $\beta$ appears only in the form $\beta t_{peak}$, changing $\beta$ is equivalent to changing the rise time. 
As was shown in \citeauthor{Khatami19} an increase in $\beta$ mimics an increase of $^{56}$Ni mixing out through the SN ejecta. Arnett's rule has been shown to be more valid for well mixed sources, corresponding to $\beta$ $\sim$1.9. Therefore, following the suggested values of $\beta$, Arnett's rule is more accurate for $^{56}$Ni mass estimations for SNe~Ic than for SNe~IIb (see Appendix \ref{sec:A-1}).\\
\indent
We now investigate how much \citeauthor{Khatami19} $^{56}$Ni masses change when different $\beta$ values are assumed.
As we want to test the robustness of our conclusion of distinct $^{56}$Ni masses between SE-SNe and SNe~II, we recalculate $^{56}$Ni masses using the lowest value of $\beta=0.82$ for all the SE~SN of our sample; i.e., that which will produce the lowest $^{56}$Ni masses for SE-SNe.
This new $^{56}$Ni mass distribution - assuming $\beta=0.82$ - is then compared to that of SNe~II and a KS test p-value of $3.4\%$ is obtained.
Therefore, while the significance of the difference in the distributions is lessened (as one would expect), the difference is still present. In addition, assuming this low $\beta=0.82$ value
produces many $^{56}$Ni masses for SE-SNe lower than the Tail method. This does not make sense, as the $^{56}$Ni masses from the tail luminosities are strict lower limits due to the non-negligible gamma ray escape fraction (see the steepness of light curve tails in Fig.~\ref{fig:allbolos}). At the same time, this result suggests that for at least some SE-SNe $^{56}$Ni is significantly mixed through the ejecta. Indeed, even using the suggested $\beta$ values, $^{56}$Ni masses from \citeauthor{Khatami19} are more or less the same as those from the Tail. This latter observation suggests that $^{56}$Ni may be even more mixed than implied by the suggested $\beta$ values.
Future work should explore ways to estimate the $^{56}$Ni mixing from observations \citep[e.g.][]{yoo19}, in order to constrain $\beta$ and provide more accurate measurements.

\subsubsection{Extinction}
\begin{figure}[ht!]
 \centering
\includegraphics[width=\linewidth]{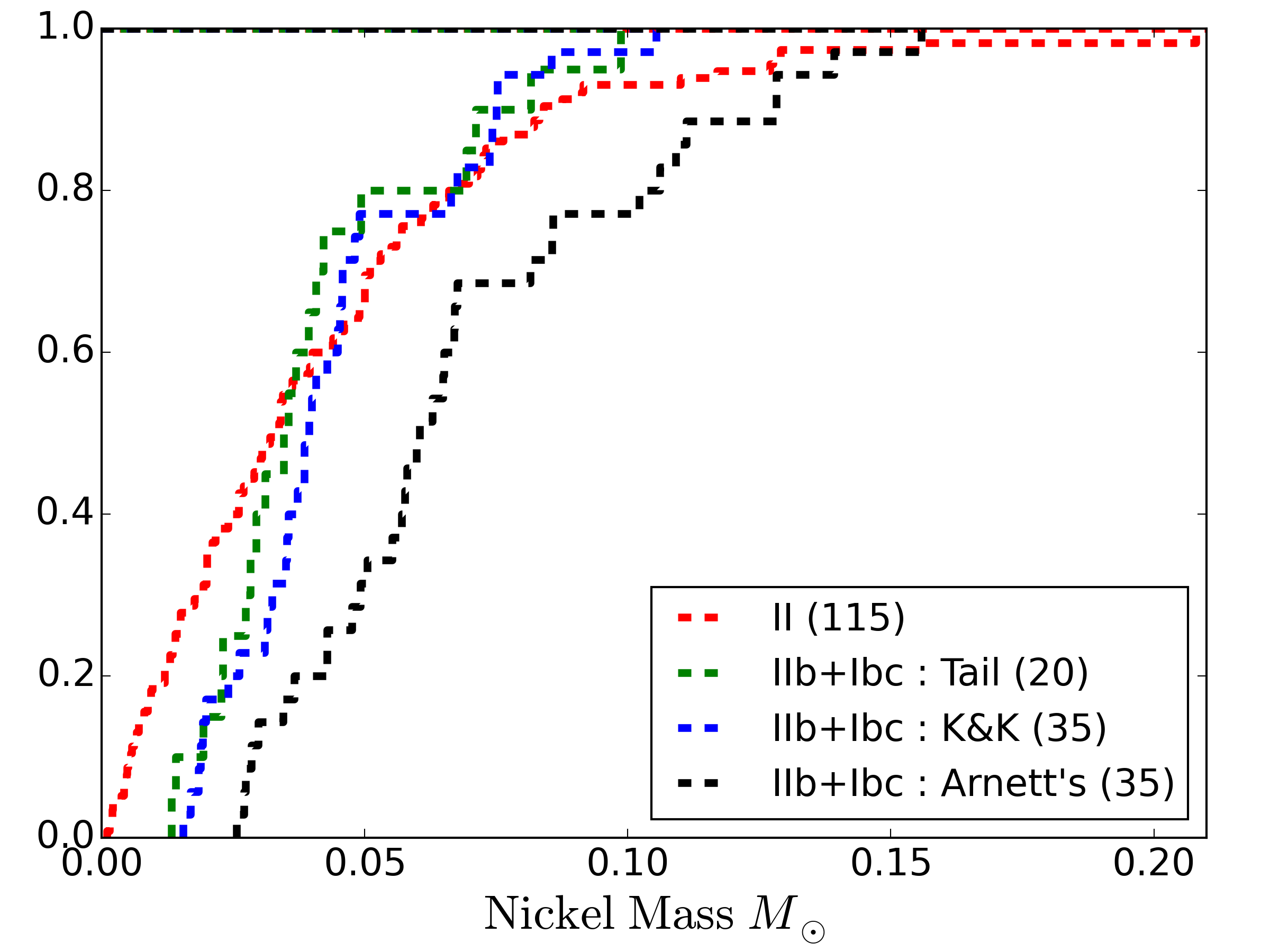}
\caption{Cumulative distributions of the SE-SN $^{56}$Ni masses derived through the three methods outlined in the text, but not corrected for host galaxy extinction, compared to that of SNe~II  \citep{and19}.\label{fig:hist_all_local}}
\end{figure}

In Fig.~\ref{fig:hist_all_local} we show again the $^{56}$Ni mass cumulative distributions (compared to that of SNe~II), but this time we only correct SE-SN photometry for Galactic reddening, and neglect host galaxy extinction corrections. 
We do this to test the how much the uncertainty of host galaxy reddening corrections affects our results.\\
\indent As expected, the cumulative distributions are now closer, with $^{56}$Ni differences reduced. When repeating the KS tests we 
find that the difference between the Tail SE-SN and SN~II (that have been corrected for host galaxy extinction) distributions is no longer statistically significant (p value of $28.6\%$). Statistically significant differences persist when using SE-SN values from Khatami ($0.022\%$) and Arnett ($<10^{-3}$~\%). Thus, there is some suggestion that uncertainties in host galaxy extinction affect our results, and could explain \textit{some} of the SE-SN--SNe~II differences. However, the test here (as in the previous two subsections) is the extreme case that is very unlikely to be true. Indeed, we have not recalculated SN~II $^{56}$Ni masses assuming zero host galaxy extinction which would have been a fairer comparison. There is no reason to believe that SE-SNe should suffer considerably less host extinction than SNe~II (see additional discussion in Section 5 of \citealt{and19}). Of course, it is clear that if we analysed SNe~II without extinction corrections, the result (of a clear difference between SN~II and SE-SN $^{56}$Ni masses) would be stronger through this test.\\ 
\indent Finally, even though the Tail $^{56}$Ni masses of SE-SNe estimated without host extinction corrections are not statistically distinct from SNe~II, the lower $^{56}$Ni mass tail of the SE-SN distribution still ends at much higher values than SNe~II (see bottom left corner of Fig.~\ref{fig:hist_all_local}).
While a deeper understanding of CC~SN host galaxy extinction is certainly warranted, we do not believe that uncertainties in this parameter are driving our results. 

\subsubsection{SN~II and SE-SN samples}
As outlined above and in \cite{and19}, the SNe we analyse in this work in no way form a complete sample. They were discovered by a large number of different surveys and data were collected by a large number of collaborations. Thus, there are many selection effects that will affect the nature of our final samples; correcting for these is not possible\footnote{One could try to assemble a volume-limited sample of CC~SNe taken from a specific survey/follow-up program, however that is beyond the scope of the current work.}. However, we can test whether biases exist in the current data set that may  produce differences in $^{56}$Ni mass between different SN types.\\
\indent Distances for our SE-SN sample are listed in Table~\ref{tab:table}. Excluding the SNe~Ic-GRB, the mean distance of this sample is 46.7 Mpc. 
The 115 SNe~II from \citeauthor{and19} have a mean distance of 42.7 Mpc. Thus there is little difference between the distances of the two samples. 
Given that SE-SN maximum-light luminosities are directly tied to the synthesised $^{56}$Ni mass, higher $^{56}$Ni mass SE-SNe will be easier to detect. Therefore, to test this bias further we split our SE-SN sample in half, using the median distance of the sample, and calculate a mean Arnett $^{56}$Ni mass for each distribution. The mean $^{56}$Ni mass of the more  distant half is 0.10\msun, while the less distant half has a mean of 0.08. Given the low number of events in each half, together with the spread in their distributions, there is no clear difference in SE-SN $^{56}$Ni mass with distance. Therefore, we conclude that there is no significant distance-selection effect causing our results. One possibility remains: that all surveys simply miss those SE-SNe exploding with very little $^{56}$Ni. We discuss this further in the next section.

\section{Discussion}
\label{sec:Discussion}
In this work we have tested the robustness of the results and conclusions presented in \cite{and19}, i.e. that the overall $^{56}$Ni distribution of SE-SNe is significantly larger than for SNe~II. 
This has been achieved through concentrating on SE-SN $^{56}$Ni masses as it was believed that these are the most uncertain. We thus defined a well-observed (in wavelength and time coverage) sample of 37 SE-SNe (14 IIb, 13 Ib, 6 Ic, 2 Ic-BL and 2 Ic-GRB), and proceeded to produce bolometric light curves and estimate $^{56}$Ni masses.\\
\indent \cite{and19} compiled literature $^{56}$Ni masses, where the vast majority of SE-SN values were estimated using `Arnett's rule' \citep{arn82}, with many cases of SNe with poorly sampled photometry or epochs included. Here, we also re-estimate $^{56}$Ni masses using Arnett's rule, but using a small sample of SE-SNe with high quality data. When doing this, we still see much higher $^{56}$Ni values for SE-SNe (with a smaller sample) than SNe~II. A key result discussed in \cite{and19} was the existence of a $^{56}$Ni tail out to values as high as $>$1\msun\ of $^{56}$Ni. The sample of well-observed SE-SNe included in the present study do not show such a tail, with a highest value of 0.21\msun. Thus the SE-SN $^{56}$Ni distribution presented here is less in conflict with explosion models (although our values are lower limits, see below for further discussion).\\ 
\indent We then use two additional methods (Section \ref{nimassest}) to estimate $^{56}$Ni masses.
The first follows \cite{Khatami19} and the second uses the tail luminosity to estimate lower-limit $^{56}$Ni masses. In both cases $^{56}$Ni mass differences between SE-SNe and SNe~II are smaller (than Arnett) but are still statistically significant.\\
\indent It is often argued that the Arnett rule overestimates $^{56}$Ni masses for many SE-SNe because of some of its limiting assumptions. This was discussed at length by \cite{des15b} and \cite{des16}. Those authors estimated (through detailed light curve modelling) that Arnett overestimates $^{56}$Ni masses by 50\%. However, lowering SE-SN $^{56}$Ni masses by this amount would not remove the SN~II--SE-SN $^{56}$Ni mass difference.\\
\indent Arnett's rule and \cite{Khatami19} both make significant assumptions in their formalisms, such as the degree of mixing and the value and time dependence of the ejecta opacity. 
In addition, both the analytic formulae used in this work and detailed light curve models in the literature assume spherical symmetry. A number of SE-SNe may arise from significantly asymmetric explosions and thus $^{56}$Ni masses determined assuming spherical symmetry will give incorrect values.
Importantly, our Tail estimates giving lower limits do not contain significant assumptions. The result that Tail $^{56}$Ni masses for SE-SNe are still significantly higher than SNe~II is thus our most robust result.
Finally, we note that detailed light curve modelling of SN~1993J (e.g. \citealt{woo94}) and SN~2011dh (e.g. \citealt{ber12}) give $^{56}$Ni masses that are reasonably in agreement with our estimates.\\
\indent In the previous section we investigated various systematics that exist in $^{56}$Ni mass estimation methodologies. While these systematics clearly affect $^{56}$Ni mass estimates, we argued that none of them are likely to be large enough (and necessarily in the correct direction) to produce the observed SE-SN--SN~II $^{56}$Ni mass distribution differences. 
In this analysis we have concentrated our efforts on SE-SNe. We assume that the SN~II $^{56}$Ni masses in the literature (and compiled by \citeauthor{and19}) are robust. This assumption is based on the majority of their late-time light curves declining at the rate predicted by $^{56}$Co (implying full trapping of the radioactive emission, in contrast to SE-SNe). Therefore, the $^{56}$Ni mass estimation follows directly from the generalised form of equation (3): t$_{peak}$ is replaced by t$_{epoch}$ that is the epoch at which one measures the luminosity during the tail (the same as for the `Tail' values for SE-SNe, but for SNe~II these are not lower limits). Thus the method for SNe~II is robust from both a theoretical and observational viewpoint.\\
\indent The SN~II method is still affected by the systematics discussed in Section \ref{sec:analysis}. With respect to uncertainties in explosion epochs, this error still propagates to an error in the $^{56}$Ni mass. However, this occurs at a much lower level due to the fractional uncertainty on t$_{epoch}$ (at the epochs when the luminosity is measured for SNe~II, $>$120 days post explosion) being much lower than the fractional uncertainty on t$_{peak}$. Host galaxy extinction estimations are just as uncertain for SNe~II as they are for SE-SNe. However, \cite{dej18} argued that apart from a few highly reddened objects most SNe~II suffer from negligible extinction and colour diversity is intrinsic to the SNe themselves. Most literature studies have assumed a higher level of host extinction for SNe~II than \citeauthor{dej18}. Thus, this systematic is more likely to push SN~II $^{56}$Ni masses to lower values: not the higher ones needed to remove the SN~II--SE-SN $^{56}$Ni difference.\\
\indent SN~II bolometric corrections rely heavily on the exquisite data of SN~1987A (see applications in e.g. \citealt{ham03b,ber09, val16}). If other SNe~II do not follow the same colour evolution as SN~1987A, the bolometric corrections employed may be in error.
In Appendix B we present a first-order analysis of this issue.
We estimate that the integrated flux in the range 3500-9000\AA \ (i.e., the range generally covered by optical followup of SNe~II) with respect to the flux in the $V$ band (the band often used to directly infer $^{56}$Ni masses) varies by $\pm0.25$ mags around the correction for SN1987A. This would translate to a dispersion of $\approx 25\%$ in SN~II $^{56}$Ni masses; however the deviations are not systematically skewed in the direction that would make SN~II values larger. Thus, while a more detailed study is warranted, we do not believe that errors in bolometric corrections are the origin of our results.\footnote{Higher $^{56}$Ni masses for SNe~II were recently published by \cite{ric19}. However, it is not clear why those authors (for the same SNe) obtain such higher values than elsewhere.}.\\
\indent Following the above summary of our work, we conclude that real, intrinsic differences exist in $^{56}$Ni masses between SE-SNe (type IIb, Ib and Ic) and SNe~II. Next, we further discuss the implications of this conclusion.

\subsection{CC~SN progenitors and explosions}
CC~SNe are spectroscopically classified into SNe~II or SE~SNe based on the detection of long-lasting broad hydrogen features in the former. That SE-SNe do not show these features leads to their naming: their outer hydrogen envelopes have been `stripped' before explosion. How progenitor stars lose this mass has long been debated. Massive stars lose material due to winds (either steady or eruptive), and the strength of these winds increases with increasing Zero Age Main Sequence (ZAMS) mass. Thus, historically SE-SNe were generally assumed to arise from more massive progenitors than SNe~II, where stellar winds are strong enough to remove the vast majority of their hydrogen envelopes and stars explode during the Wolf-Rayet phase (see e.g. \citealt{beg86}). Through this scenario the progenitors of SE-SNe have ZAMS masses higher than 25\msun, thus being significantly higher mass than SNe~II\footnote{Direct progenitor detections constrain SN~II progenitors to be between 8 and 20\msun\ \citep{sma15b}.}.\\
\indent Alternatively, the mass stripping may occur through binary interaction (e.g. \citealt{pod92}). In this scenario SE-SN progenitors have ZAMS masses in a very similar mass range to SN~II, with the presence of a close binary companion being the factor that produces distinct SN types.\\
\indent During the last decade evidence has mounted in favour of the low-mass binary scenario, at least for the majority of SE-SNe. From analysis of the width of bolometric light curves around peak luminosity, a number of investigations have concluded that SE-SN ejecta masses are relatively low, implying low-mass (i.e. similar mass to SNe~II) ZAMS masses (e.g. \citealt{dro11,tad15,lym16,tad18a,pre19})\footnote{However, these analyses use the same `Arnett' formalism used for $^{56}$Ni mass estimates. 
Given the clear uncertainty in this methodology (i.e. the differences between Arnett and other $^{56}$Ni values discussed in this paper) these ejecta masses may also need to be taken with caution.}. In addition, the relatively high rates of SE-SNe have been claimed to be incompatible with arising from only $>$25\msun\ progenitors \citep{smi11b}.
While for SNe~Ib and Ic there exist very few progenitor detections on pre-explosion images, there exist a small number of direct detections for SNe~IIb. In all cases these progenitors are constrained to be less than 20\msun\ ZAMS \citep{mau09,van13,tar17}.
At the same time, studies of the local environments of SNe~Ic within host galaxies suggests that these events arise from shorter lived, and therefore more massive progenitors than the rest of the CC~SN population (see e.g. \citealt{and12,kan17,kun18,gal18,mau18}).
However, the general current consensus is that at least a significant majority of SE-SN progenitors arise from binary stars with ZAMS masses similar to SNe~II (see review in \citealt{smi14c}, for a detailed discussion on this topic).\\
\indent Following the above, if SNe~II and SE-SNe arise from similar mass progenitors then how can the $^{56}$Ni mass differences presented here be explained? Similar mass progenitors will produce similar core structures (for the majority of events), whether the progenitor is a single star or part of a binary system.
Thus, there appears to be an inconsistency between the similar-mass progenitors between different CC~SNe as generally discussed in the literature, and the clear $^{56}$Ni mass differences discussed in this paper. Further advances in our understanding of the underlying physics of different CC~SN progenitors and their explosions, together with the effects of binary interaction are clearly required.\\

The standard explosion mechanism for CC~SNe is that of neutrino driven explosions (see \citealt{mul16} for a recent review). As discussed in detail in \cite{and19}, several works have estimated nucleosynthetic yields within this explosion framework (e.g.~\citealt{ugl12,pej15c,suk16,suw19,cur19}).
The high $^{56}$Ni masses compiled for SE-SNe by \citeauthor{and19} were seen to be inconsistent with the $^{56}$Ni values from explosion models. Here (as outlined above), for our smaller sample of well-observed SE-SNe we no longer obtain $^{56}$Ni masses in excess of 0.2\msun, even for the Arnett values (although we again emphasise that $^{56}$Ni masses we present in this article are lower limits due to the lack of observations outside the $B$ and $H$ bands). Thus, the degree of inconsistency between SE-SN $^{56}$Ni masses and those predicted by neutrino-driven explosions is lowered. 
$^{56}$Ni masses produced by explosion models are generally predicted to increase with increasing ZAMS mass. When compared to explosion model predictions the higher 
$^{56}$Ni masses for SE-SNe than SNe~II suggests higher ZAMS masses for the latter. Indeed, given that our presented $^{56}$Ni values are lower limits, a significant fraction of the masses presented here are at the high end of explosion model predictions, possibly suggesting $>$20\msun\ progenitor masses for a significant number of SE-SNe, in contradiction with the discussion above on the consensus of lower-mass binary progenitors for the majority of SE-SNe.\\
\indent Recently, \cite{ert19} compared their model light curves and $^{56}$Ni mass predictions to a large number of published SE-SNe light curves (from e.g. \citealt{lym16,pre19}). These authors were not able to reproduce the observed light curves and $^{56}$Ni estimates for a large fraction of the literature samples, through standard neutrino-driven explosions. Thus, they concluded that an additional power source is required for such SE-SNe (possibilities include a magnetar or circumstellar interaction). If an additional power source is indeed present, this would lower the required $^{56}$Ni masses to power the light curves and thus SE-SN $^{56}$Ni masses could become closer to those of SNe~II\footnote{However, one could then naturally ask: why does such an additional power source exist for SE-SNe and not SNe~II?}.\\
\indent From the observational side, there remains the possibility that there is a family of low $^{56}$Ni, intrinsically dim objects that have been missed by SN searches. Indeed, (as discussed previously) the remaining $^{56}$Ni mass difference between SE-SNe and SNe~II presented in this work seems to be strongly driven by a lack of low SE-SN $^{56}$Ni masses (see Fig.~\ref{fig:hist_all}). If we remove SN~II values below the minimum SE-SN $^{56}$Ni mass (0.015M$_\odot$, for SN~2017czd) the distributions become statistically consistent, with KS p-values greater than 40\% (except for Arnett's values that remain discrepant; p-value below $10^{-3}$). However, we once again reiterate that our $^{56}$Ni values are strict lower limits. If we arbitrarily increase our SE-SN $^{56}$Ni values by 20\% (roughly accounting for the missing flux outside our filter range), a small tension remains; between the Khatami SE-SN and SN~II distributions a KS test gives a p-value of 9\%.\\
\indent There is now significant evidence that the majority of massive stars 
are found in binary systems where their orbital periods are such that significant interaction with the companion star will occur during the star's life (e.g. \citealt{san12_2}). There is also some indication of an increasing relative fraction of short period systems in late-B and O-type stars \citep{moe17}. 
Such observations imply that some SE-SNe may originate from quite low-mass massive stars. At the lowest mass range (i.e. the lowest masses from which stars will still explode as CC~SNe) it is possible that such progenitors explode with small ejecta masses and little $^{56}$Ni.\\
\indent This lack of low-mass $^{56}$Ni SE-SNe in current samples begs the question: if a SE-SN were to explode with a $^{56}$Ni mass of 0.001\msun\ (the extreme lower end of the SN~II distribution, \citealt{mul17,and19}) how dim and fast evolving would such an event be and would current surveys detect such explosions?\\
\indent Scenarios for these type of objects exist in theory \citep{kle14,tau15}, and a few transients, associated with the family of ``Calcium gap'' events and potential precursor systems of binary neutron star mergers, have been observed recently \citep{dek18,dek18b,she19}. The latter class, named ultra-stripped SNe, arise from tight (period $<2$ days) binary sytems of a Helium star and a compact object (such as a neutron star). The pre-SN star would contain less than 0.2M$_\odot$ of helium in the envelope and the final explosion would eject $\approx 0.1$\msun , with a $^{56}$Ni mass of $\approx 0.01$\msun\ (see e.g. \citealt{moriya17}).
Ultra-stripped SNe are expected to have a very fast evolution (rise time of less than 10 days) and to be very dim - with the exception of progenitors with a pre-SN extended progenitor or when the SN ejecta interacts with CSM \citep{kle14,kle18a}. The rate for ultra-stripped SNe is expected to be 1\%\ of all CCSNe, and therefore their inclusion in the current study is unlikely to significantly reduce the SN~II--SE-SN $^{56}$Ni mass difference. Future modelling of low-mass binary systems producing low-$^{56}$Ni mass CC~SNe is certainly warranted.

\label{sec:discussion}

\section{Conclusions}
The amount of radioactive material synthesised in a SN explosion is a fundamental parameter that determines the transient behaviour of all SN types. The mass of $^{56}$Ni produced in CC~SNe is determined by the core-structure at the explosion epoch, together with the explosion energy. Therefore constraining $^{56}$Ni masses for different CC~SN types can shed light on differences in progenitors and explosion properties.\\
\indent We conclude that real intrinsic differences in $^{56}$Ni mass exist between observed SE-SNe and SNe~II, differences that persist when different systematic errors in $^{56}$Ni mass estimations are analysed. 
In particular, a $^{56}$Ni mass difference is still observed when we use the radioactive tail luminosities to obtain $^{56}$Ni mass lower limits. This Tail methodology is extremely robust and we suggest that effort is made to obtain additional well-sampled multi-colour late time observations of SE-SNe. The $^{56}$Ni discrepancy we present in this work is driven by a lack of low $^{56}$Ni mass SE-SNe.\\
\indent That SE-SNe are observed to produce larger $^{56}$Ni masses than SNe~II implies significant differences in their progenitor properties and/or explosion mechanisms. A full understanding of which parameter produces these $^{56}$Ni mass differences is of utmost importance for our understanding of massive star explosions.

\label{sec:conclusions}

\begin{acknowledgements}
This work was funded by ESO, as part of a short term intership at ESO-Chile at Vitacura, during the first quarter of 2019. The authors acknowledge fruitful discussions and comments from Luc Dessart and Jose Lu\'is Prieto. 
\end{acknowledgements}

%
%
\bibliographystyle{aa}
\bibliography{Reference}

\begin{thebibliography}{87}
\expandafter\ifx\csname natexlab\endcsname\relax\def\natexlab#1{#1}\fi

\bibitem[{{Anderson}(2019)}]{and19}
{Anderson}, J.~P. 2019, \aap, 628, A7

\bibitem[{{Anderson} {et~al.}(2012){Anderson}, {Habergham}, {James}, \&
  {Hamuy}}]{and12}
{Anderson}, J.~P., {Habergham}, S.~M., {James}, P.~A., \& {Hamuy}, M. 2012,
  \mnras, 424, 1372

\bibitem[{{Anderson} {et~al.}(2014)}]{and14a}
{Anderson}, J.~P. {et~al.} 2014, \apj, 786, 67

\bibitem[{{Arcavi} {et~al.}(2011){Arcavi}, {Gal-Yam}, {Yaron}, {Sternberg},
  {Rabinak}, {Waxman}, {Kasliwal}, {Quimby}, {Ofek}, {Horesh}, {Kulkarni},
  {Filippenko}, {Silverman}, {Cenko}, {Li}, {Bloom}, {Sullivan}, {Nugent},
  {Poznanski}, {Gorbikov}, {Fulton}, {Howell}, {Bersier}, {Riou},
  {Lamotte-Bailey}, {Griga}, {Cohen}, {Hachinger}, {Polishook}, {Xu},
  {Ben-Ami}, {Manulis}, {Walker}, {Maguire}, {Pan}, {Matheson}, {Mazzali},
  {Pian}, {Fox}, {Gehrels}, {Law}, {James}, {Marchant}, {Smith}, {Mottram},
  {Barnsley}, {Kandrashoff}, \& {Clubb}}]{arc11}
{Arcavi}, I., {Gal-Yam}, A., {Yaron}, O., {et~al.} 2011, \apjl, 742, L18

\bibitem[{{Arnett}(1982)}]{arn82}
{Arnett}, W.~D. 1982, \apj, 253, 785

\bibitem[{{Arnett} \& {Clayton}(1970)}]{arn70}
{Arnett}, W.~D. \& {Clayton}, D.~D. 1970, \nat, 227, 780

\bibitem[{{Begelman} \& {Sarazin}(1986)}]{beg86}
{Begelman}, M.~C. \& {Sarazin}, C.~L. 1986, \apj, 302, L59

\bibitem[{{Bersten} \& {Hamuy}(2009)}]{ber09}
{Bersten}, M.~C. \& {Hamuy}, M. 2009, \apj, 701, 200

\bibitem[{{Bersten} {et~al.}(2012)}]{ber12}
{Bersten}, M.~C. {et~al.} 2012, \apj, 757, 31

\bibitem[{{Bianco} {et~al.}(2014)}]{bia14}
{Bianco}, F.~B. {et~al.} 2014, \apjs, 213, 19

\bibitem[{{Bufano} {et~al.}(2014){Bufano}, {Pignata}, {Bersten}, {Mazzali},
  {Ryder}, {Margutti}, {Milisavljevic}, {Morelli}, {Benetti}, {Cappellaro},
  {Gonzalez-Gaitan}, {Romero-Ca{\~n}izales}, {Stritzinger}, {Walker},
  {Anderson}, {Contreras}, {de Jaeger}, {F{\"o}rster}, {Gutierrez}, {Hamuy},
  {Hsiao}, {Morrell}, {Olivares E.}, {Paillas}, {Parker}, {Pian}, {Pickering},
  {Sanders}, {Stockdale}, {Turatto}, {Valenti}, {Fesen}, {Maza}, {Nomoto},
  {Phillips}, \& {Soderberg}}]{buf14}
{Bufano}, F., {Pignata}, G., {Bersten}, M., {et~al.} 2014, \mnras, 439, 1807

\bibitem[{{Cano} {et~al.}(2011){Cano}, {Bersier}, {Guidorzi}, {Kobayashi},
  {Levan}, {Tanvir}, {Wiersema}, {D'Avanzo}, {Fruchter}, {Garnavich}, {Gomboc},
  {Gorosabel}, {Kasen}, {Kopa{\v{c}}}, {Margutti}, {Mazzali}, {Melandri},
  {Mundell}, {Nugent}, {Pian}, {Smith}, {Steele}, {Wijers}, \&
  {Woosley}}]{can11}
{Cano}, Z., {Bersier}, D., {Guidorzi}, C., {et~al.} 2011, \apj, 740, 41

\bibitem[{Chieffi \& Limongi(2017)}]{chi17}
Chieffi, A. \& Limongi, M. 2017, The Astrophysical Journal, 836, 79

\bibitem[{{Colgate} \& {McKee}(1969)}]{col69}
{Colgate}, S.~A. \& {McKee}, C. 1969, \apj, 157, 623

\bibitem[{{Curtis} {et~al.}(2019){Curtis}, {Ebinger}, {Fr{\"o}hlich}, {Hempel},
  {Perego}, {Liebend{\"o}rfer}, \& {Thielemann}}]{cur19}
{Curtis}, S., {Ebinger}, K., {Fr{\"o}hlich}, C., {et~al.} 2019, \apj, 870, 2

\bibitem[{{De} {et~al.}(2018{\natexlab{a}}){De}, {Kasliwal}, {Cantwell}, {Cao},
  {Cenko}, {Gal-Yam}, {Johansson}, {Kong}, {Kulkarni}, {Lunnan}, {Masci},
  {Matuszewski}, {Mooley}, {Neill}, {Nugent}, {Ofek}, {Perrott},
  {Rebbapragada}, {Rubin}, {O' Sullivan}, \& {Yaron}}]{dek18}
{De}, K., {Kasliwal}, M.~M., {Cantwell}, T., {et~al.} 2018{\natexlab{a}}, \apj,
  866, 72

\bibitem[{{De} {et~al.}(2018{\natexlab{b}}){De}, {Kasliwal}, {Ofek}, {Moriya},
  {Burke}, {Cao}, {Cenko}, {Doran}, {Duggan}, {Fender}, {Fransson}, {Gal-Yam},
  {Horesh}, {Kulkarni}, {Laher}, {Lunnan}, {Manulis}, {Masci}, {Mazzali},
  {Nugent}, {Perley}, {Petrushevska}, {Piro}, {Rumsey}, {Sollerman},
  {Sullivan}, \& {Taddia}}]{dek18b}
{De}, K., {Kasliwal}, M.~M., {Ofek}, E.~O., {et~al.} 2018{\natexlab{b}},
  Science, 362, 201

\bibitem[{{de Jaeger} {et~al.}(2018){de Jaeger}, {Anderson}, {Galbany},
  {Gonz{\'a }lez-Gait{\'a}n}, {Hamuy}, {Phillips}, {Stritzinger}, {Contreras},
  {Folatelli}, {Guti{\'e}rrez}, {Hsiao}, {Morrell}, {Suntzeff}, {Dessart}, \&
  {Filippenko}}]{dej18}
{de Jaeger}, T., {Anderson}, J.~P., {Galbany}, L., {et~al.} 2018, \mnras, 476,
  4592

\bibitem[{{Dessart} {et~al.}(2011){Dessart}, {Hillier}, {Livne}, {Yoon},
  {Woosley}, {Waldman}, \& {Langer}}]{des11}
{Dessart}, L., {Hillier}, D.~J., {Livne}, E., {et~al.} 2011, \mnras, 414, 2985

\bibitem[{{Dessart} {et~al.}(2013){Dessart}, {Hillier}, {Waldman}, \&
  {Livne}}]{des13}
{Dessart}, L., {Hillier}, D.~J., {Waldman}, R., \& {Livne}, E. 2013, \mnras,
  433, 1745

\bibitem[{{Dessart} {et~al.}(2015){Dessart}, {Hillier}, {Woosley}, {Livne},
  {Waldman}, {Yoon}, \& {Langer}}]{des15b}
{Dessart}, L., {Hillier}, D.~J., {Woosley}, S., {et~al.} 2015, \mnras, 453,
  2189

\bibitem[{{Dessart} {et~al.}(2016){Dessart}, {Hillier}, {Woosley}, {Livne},
  {Waldman}, {Yoon}, \& {Langer}}]{des16}
{Dessart}, L., {Hillier}, D.~J., {Woosley}, S., {et~al.} 2016, \mnras, 458,
  1618

\bibitem[{{Drout} {et~al.}(2011)}]{dro11}
{Drout}, M.~R. {et~al.} 2011, \apj, 741, 97

\bibitem[{{Ensman} \& {Woosley}(1988)}]{ens88}
{Ensman}, L.~M. \& {Woosley}, S.~E. 1988, \apj, 333, 754

\bibitem[{{Ergon} {et~al.}(2015){Ergon}, {Jerkstrand}, {Sollerman},
  {Elias-Rosa}, {Fransson}, {Fraser}, {Pastorello}, {Kotak}, {Taubenberger},
  {Tomasella}, {Valenti}, {Benetti}, {Helou}, {Kasliwal}, {Maund}, {Smartt}, \&
  {Spyromilio}}]{erg15}
{Ergon}, M., {Jerkstrand}, A., {Sollerman}, J., {et~al.} 2015, \aap, 580, A142

\bibitem[{{Ertl} {et~al.}(2019){Ertl}, {Woosley}, {Sukhbold}, \&
  {Janka}}]{ert19}
{Ertl}, T., {Woosley}, S.~E., {Sukhbold}, T., \& {Janka}, H.~T. 2019, arXiv
  e-prints, arXiv:1910.01641

\bibitem[{{Foley} {et~al.}(2014){Foley}, {McCully}, {Jha}, {Bildsten}, {Fong},
  {Narayan}, {Rest}, \& {Stritzinger}}]{fol14}
{Foley}, R.~J., {McCully}, C., {Jha}, S.~W., {et~al.} 2014, \apj, 792, 29

\bibitem[{{Gal-Yam}(2017)}]{gal17}
{Gal-Yam}, A. 2017, {Observational and Physical Classification of Supernovae},
  195

\bibitem[{{Galbany} {et~al.}(2018){Galbany}, {Anderson}, {S{\'a}nchez},
  {Kuncarayakti}, {Pedraz}, {Gonz{\'a}lez-Gait{\'a}n}, {Stanishev},
  {Dom{\'\i}nguez}, {Moreno-Raya}, {Wood-Vasey}, {Mour{\~a}o}, {Ponder},
  {Badenes}, {Moll{\'a}}, {L{\'o}pez-S{\'a}nchez}, {Rosales-Ortega},
  {V{\'\i}lchez}, {Garc{\'\i}a-Benito}, \& {Marino}}]{gal18}
{Galbany}, L., {Anderson}, J.~P., {S{\'a}nchez}, S.~F., {et~al.} 2018, \apj,
  855, 107

\bibitem[{{Guillochon} {et~al.}(2017){Guillochon}, {Parrent}, {Kelley}, \&
  {Margutti}}]{gui17}
{Guillochon}, J., {Parrent}, J., {Kelley}, L.~Z., \& {Margutti}, R. 2017, \apj,
  835, 64

\bibitem[{{Hamuy} {et~al.}(2003)}]{ham03b}
{Hamuy}, M. {et~al.} 2003, \nat, 424, 651

\bibitem[{{Hunter} {et~al.}(2009){Hunter}, {Valenti}, {Kotak}, {Meikle},
  {Taubenberger}, {Pastorello}, {Benetti}, {Stanishev}, {Smartt}, {Trundle},
  {Arkharov}, {Bufano}, {Cappellaro}, {Di Carlo}, {Dolci}, {Elias-Rosa},
  {Frandsen}, {Fynbo}, {Hopp}, {Larionov}, {Laursen}, {Mazzali}, {Navasardyan},
  {Ries}, {Riffeser}, {Rizzi}, {Tsvetkov}, {Turatto}, \& {Wilke}}]{hun09}
{Hunter}, D.~J., {Valenti}, S., {Kotak}, R., {et~al.} 2009, \aap, 508, 371

\bibitem[{{Jerkstrand} {et~al.}(2012){Jerkstrand}, {Fransson}, {Maguire},
  {Smartt}, {Ergon}, \& {Spyromilio}}]{jer12}
{Jerkstrand}, A., {Fransson}, C., {Maguire}, K., {et~al.} 2012, \aap, 546, A28

\bibitem[{{Kangas} {et~al.}(2017){Kangas}, {Portinari}, {Mattila}, {Fraser},
  {Kankare}, {Izzard}, {James}, {Gonz{\'a}lez-Fern{\'a}ndez}, {Maund}, \&
  {Thompson}}]{kan17}
{Kangas}, T., {Portinari}, L., {Mattila}, S., {et~al.} 2017, \aap, 597, A92

\bibitem[{{Khatami} \& {Kasen}(2018)}]{kha18}
{Khatami}, D.~K. \& {Kasen}, D.~N. 2018, arXiv e-prints, arXiv:1812.06522

\bibitem[{{Khatami} \& {Kasen}(2019)}]{Khatami19}
{Khatami}, D.~K. \& {Kasen}, D.~N. 2019, \apj, 878, 56

\bibitem[{{Kleiser} \& {Kasen}(2014)}]{kle14}
{Kleiser}, I. K.~W. \& {Kasen}, D. 2014, \mnras, 438, 318

\bibitem[{{Kleiser} {et~al.}(2018){Kleiser}, {Kasen}, \& {Duffell}}]{kle18a}
{Kleiser}, I. K.~W., {Kasen}, D., \& {Duffell}, P.~C. 2018, \mnras, 475, 3152

\bibitem[{{Kuncarayakti} {et~al.}(2018){Kuncarayakti}, {Anderson}, {Galbany},
  {Maeda}, {Hamuy}, {Aldering}, {Arimoto}, {Doi}, {Morokuma}, \&
  {Usuda}}]{kun18}
{Kuncarayakti}, H., {Anderson}, J.~P., {Galbany}, L., {et~al.} 2018, \aap, 613,
  A35

\bibitem[{{L{\"u}} {et~al.}(2018){L{\"u}}, {Lan}, {Zhang}, {Liang}, {Kann},
  {Du}, \& {Shen}}]{Lu2018}
{L{\"u}}, H.-J., {Lan}, L., {Zhang}, B., {et~al.} 2018, \apj, 862, 130

\bibitem[{{Lyman} {et~al.}(2016){Lyman}, {Bersier}, {James}, {Mazzali},
  {Eldridge}, {Fraser}, \& {Pian}}]{lym16}
{Lyman}, J.~D., {Bersier}, D., {James}, P.~A., {et~al.} 2016, \mnras, 457, 328

\bibitem[{{Maeda} {et~al.}(2008){Maeda}, {Kawabata}, {Mazzali}, {Tanaka},
  {Valenti}, {Nomoto}, {Hattori}, {Deng}, {Pian}, {Taubenberger}, {Iye},
  {Matheson}, {Filippenko}, {Aoki}, {Kosugi}, {Ohyama}, {Sasaki}, \&
  {Takata}}]{mae08}
{Maeda}, K., {Kawabata}, K., {Mazzali}, P.~A., {et~al.} 2008, Science, 319,
  1220

\bibitem[{{Maund}(2018)}]{mau18}
{Maund}, J.~R. 2018, \mnras, 476, 2629

\bibitem[{{Maund} \& {Smartt}(2009)}]{mau09}
{Maund}, J.~R. \& {Smartt}, S.~J. 2009, Science, 324, 486

\bibitem[{{Mazzali} {et~al.}(2008){Mazzali}, {Valenti}, {Della Valle},
  {Chincarini}, {Sauer}, {Benetti}, {Pian}, {Piran}, {D'Elia}, {Elias-Rosa},
  {Margutti}, {Pasotti}, {Antonelli}, {Bufano}, {Campana}, {Cappellaro},
  {Covino}, {D'Avanzo}, {Fiore}, {Fugazza}, {Gilmozzi}, {Hunter}, {Maguire},
  {Maiorano}, {Marziani}, {Masetti}, {Mirabel}, {Navasardyan}, {Nomoto},
  {Palazzi}, {Pastorello}, {Panagia}, {Pellizza}, {Sari}, {Smartt},
  {Tagliaferri}, {Tanaka}, {Taubenberger}, {Tominaga}, {Trundle}, \&
  {Turatto}}]{maz08}
{Mazzali}, P.~A., {Valenti}, S., {Della Valle}, M., {et~al.} 2008, Science,
  321, 1185

\bibitem[{{Mirabal} {et~al.}(2006){Mirabal}, {Halpern}, {An}, {Thorstensen}, \&
  {Terndrup}}]{mir06}
{Mirabal}, N., {Halpern}, J.~P., {An}, D., {Thorstensen}, J.~R., \& {Terndrup},
  D.~M. 2006, \apjl, 643, L99

\bibitem[{{Moe} \& {Di Stefano}(2017)}]{moe17}
{Moe}, M. \& {Di Stefano}, R. 2017, \apjs, 230, 15

\bibitem[{{Morales-Garoffolo} {et~al.}(2015){Morales-Garoffolo}, {Elias-Rosa},
  {Bersten}, {Jerkstrand}, {Taubenberger}, {Benetti}, {Cappellaro}, {Kotak},
  {Pastorello}, {Bufano}, {Dom{\'\i}nguez}, {Ergon}, {Fraser}, {Gao},
  {Garc{\'\i}a}, {Howell}, {Isern}, {Smartt}, {Tomasella}, \&
  {Valenti}}]{mor15}
{Morales-Garoffolo}, A., {Elias-Rosa}, N., {Bersten}, M., {et~al.} 2015,
  \mnras, 454, 95

\bibitem[{{Moriya} {et~al.}(2017){Moriya}, {Mazzali}, {Tominaga}, {Hachinger},
  {Blinnikov}, {Tauris}, {Takahashi}, {Tanaka}, {Langer}, \&
  {Podsiadlowski}}]{moriya17}
{Moriya}, T.~J., {Mazzali}, P.~A., {Tominaga}, N., {et~al.} 2017, \mnras, 466,
  2085

\bibitem[{{M{\"u}ller}(2016)}]{mul16}
{M{\"u}ller}, B. 2016, \pasa, 33, e048

\bibitem[{{M{\"u}ller} {et~al.}(2017){M{\"u}ller}, {Prieto}, {Pejcha}, \&
  {Clocchiatti}}]{mul17}
{M{\"u}ller}, T., {Prieto}, J.~L., {Pejcha}, O., \& {Clocchiatti}, A. 2017,
  \apj, 841, 127

\bibitem[{{Nakaoka} {et~al.}(2019){Nakaoka}, {Moriya}, {Tanaka}, {Yamanaka},
  {Kawabata}, {Maeda}, {Kawabata}, {Kawahara}, {Itagaki}, {Ouchi}, {Blinnikov},
  {Tominaga}, \& {Uemura}}]{nak19}
{Nakaoka}, T., {Moriya}, T.~J., {Tanaka}, M., {et~al.} 2019, \apj, 875, 76

\bibitem[{{Olivares} {et~al.}(2012){Olivares}, {Greiner}, {Schady}, {Rau},
  {Klose}, \& {Kr{\"u}hler}}]{oli12}
{Olivares}, E.~F., {Greiner}, J., {Schady}, P., {et~al.} 2012, in IAU
  Symposium, Vol. 279, Death of Massive Stars: Supernovae and Gamma-Ray Bursts,
  ed. P.~{Roming}, N.~{Kawai}, \& E.~{Pian}, 375--376

\bibitem[{{Pastorello} {et~al.}(2008)}]{pas08}
{Pastorello}, A. {et~al.} 2008, \mnras, 389, 113

\bibitem[{{Pejcha} \& {Thompson}(2015)}]{pej15c}
{Pejcha}, O. \& {Thompson}, T.~A. 2015, \apj, 801, 90

\bibitem[{{Pignata} {et~al.}(2011){Pignata}, {Stritzinger}, {Soderberg},
  {Mazzali}, {Phillips}, {Morrell}, {Anderson}, {Boldt}, {Campillay},
  {Contreras}, {Folatelli}, {F{\"o}rster}, {Gonz{\'a}lez}, {Hamuy},
  {Krzeminski}, {Maza}, {Roth}, {Salgado}, {Levesque}, {Rest}, {Crain},
  {Foster}, {Haislip}, {Ivarsen}, {LaCluyze}, {Nysewander}, \&
  {Reichart}}]{pig11}
{Pignata}, G., {Stritzinger}, M., {Soderberg}, A., {et~al.} 2011, \apj, 728, 14

\bibitem[{{Podsiadlowski} {et~al.}(1992){Podsiadlowski}, {Joss}, \&
  {Hsu}}]{pod92}
{Podsiadlowski}, P., {Joss}, P.~C., \& {Hsu}, J.~J.~L. 1992, \apj, 391, 246

\bibitem[{{Prentice} {et~al.}(2019){Prentice}, {Ashall}, {James}, {Short},
  {Mazzali}, {Bersier}, {Crowther}, {Barbarino}, {Chen}, {Copperwheat},
  {Darnley}, {Denneau}, {Elias-Rosa}, {Fraser}, {Galbany}, {Gal-Yam},
  {Harmanen}, {Howell}, {Hosseinzadeh}, {Inserra}, {Kankare}, {Karamehmetoglu},
  {Lamb}, {Limongi}, {Maguire}, {McCully}, {Olivares E}, {Piascik}, {Pignata},
  {Reichart}, {Rest}, {Reynolds}, {Rodr{\'\i}guez}, {Saario}, {Schulze},
  {Smartt}, {Smith}, {Sollerman}, {Stalder}, {Sullivan}, {Taddia}, {Valenti},
  {Vergani}, {Williams}, \& {Young}}]{pre19}
{Prentice}, S.~J., {Ashall}, C., {James}, P.~A., {et~al.} 2019, \mnras, 485,
  1559

\bibitem[{{Prentice} {et~al.}(2018){Prentice}, {Ashall}, {Mazzali}, {Zhang},
  {James}, {Wang}, {Vink{\'o}}, {Percival}, {Short}, {Piascik}, {Huang}, {Mo},
  {Rui}, {Wang}, {Xiang}, {Xin}, {Yi}, {Yu}, {Zhai}, {Zhang}, {Hosseinzadeh},
  {Howell}, {McCully}, {Valenti}, {Cseh}, {Hanyecz}, {Kriskovics}, {P{\'a}l},
  {S{\'a}rneczky}, {S{\'o}dor}, {Szak{\'a}ts}, {Sz{\'e}kely},
  {Varga-Vereb{\'e}lyi}, {Vida}, {Bradac}, {Reichart}, {Sand}, \&
  {Tartaglia}}]{pre18b}
{Prentice}, S.~J., {Ashall}, C., {Mazzali}, P.~A., {et~al.} 2018, \mnras, 478,
  4162

\bibitem[{{Richmond} {et~al.}(1994){Richmond}, {Treffers}, {Filippenko},
  {Paik}, {Leibundgut}, {Schulman}, \& {Cox}}]{ric94}
{Richmond}, M.~W., {Treffers}, R.~R., {Filippenko}, A.~V., {et~al.} 1994, \aj,
  107, 1022

\bibitem[{{Ricks} \& {Dwarkadas}(2019)}]{ric19}
{Ricks}, W. \& {Dwarkadas}, V.~V. 2019, arXiv e-prints, arXiv:1906.07311

\bibitem[{{Roy} {et~al.}(2013){Roy}, {Kumar}, {Maund}, {Schady}, {Olivares},
  {Malesani}, {Leloudas}, {Nandi}, {Tanvir}, {Milisavljevic}, {Hjorth},
  {Misra}, {Kumar}, {Pandey}, {Sagar}, \& {Chand ola}}]{roy13}
{Roy}, R., {Kumar}, B., {Maund}, J.~R., {et~al.} 2013, \mnras, 434, 2032

\bibitem[{{Sahu} {et~al.}(2013){Sahu}, {Anupama}, \& {Chakradhari}}]{sah13}
{Sahu}, D.~K., {Anupama}, G.~C., \& {Chakradhari}, N.~K. 2013, \mnras, 433, 2

\bibitem[{{Sana} {et~al.}(2012)}]{san12_2}
{Sana}, H. {et~al.} 2012, Science, 337, 444

\bibitem[{{Schlafly} \& {Finkbeiner}(2011)}]{sch11}
{Schlafly}, E.~F. \& {Finkbeiner}, D.~P. 2011, \apj, 737, 103

\bibitem[{{Shen} {et~al.}(2019){Shen}, {Quataert}, \& {Pakmor}}]{she19}
{Shen}, K.~J., {Quataert}, E., \& {Pakmor}, R. 2019, arXiv e-prints,
  arXiv:1908.08056

\bibitem[{{Smartt} {et~al.}(2015){Smartt}, {Valenti}, {Fraser}, {Inserra},
  {Young}, {Sullivan}, {Pastorello}, {Benetti}, {Gal-Yam}, {Knapic},
  {Molinaro}, {Smareglia}, {Smith}, {Taubenberger}, {Yaron}, {Anderson},
  {Ashall}, {Balland}, {Baltay}, {Barbarino}, {Bauer}, {Baumont}, {Bersier},
  {Blagorodnova}, {Bongard}, {Botticella}, {Bufano}, {Bulla}, {Cappellaro},
  {Campbell}, {Cellier-Holzem}, {Chen}, {Childress}, {Clocchiatti},
  {Contreras}, {Dall'Ora}, {Danziger}, {de Jaeger}, {De Cia}, {Della Valle},
  {Dennefeld}, {Elias-Rosa}, {Elman}, {Feindt}, {Fleury}, {Gall},
  {Gonzalez-Gaitan}, {Galbany}, {Morales Garoffolo}, {Greggio}, {Guillou},
  {Hachinger}, {Hadjiyska}, {Hage}, {Hillebrandt}, {Hodgkin}, {Hsiao}, {James},
  {Jerkstrand}, {Kangas}, {Kankare}, {Kotak}, {Kromer}, {Kuncarayakti},
  {Leloudas}, {Lundqvist}, {Lyman}, {Hook}, {Maguire}, {Manulis}, {Margheim},
  {Mattila}, {Maund}, {Mazzali}, {McCrum}, {McKinnon}, {Moreno-Raya},
  {Nicholl}, {Nugent}, {Pain}, {Pignata}, {Phillips}, {Polshaw}, {Pumo},
  {Rabinowitz}, {Reilly}, {Romero-Ca{\~n}izales}, {Scalzo}, {Schmidt},
  {Schulze}, {Sim}, {Sollerman}, {Taddia}, {Tartaglia}, {Terreran},
  {Tomasella}, {Turatto}, {Walker}, {Walton}, {Wyrzykowski}, {Yuan}, \&
  {Zampieri}}]{sma15b}
{Smartt}, S.~J., {Valenti}, S., {Fraser}, M., {et~al.} 2015, \aap, 579, A40

\bibitem[{{Smith}(2014)}]{smi14c}
{Smith}, N. 2014, \araa, 52, 487

\bibitem[{{Smith} {et~al.}(2011){Smith}, {Li}, {Filippenko}, \&
  {Chornock}}]{smi11b}
{Smith}, N., {Li}, W., {Filippenko}, A.~V., \& {Chornock}, R. 2011, \mnras,
  412, 1522

\bibitem[{{Stritzinger} {et~al.}(2009){Stritzinger}, {Mazzali}, {Phillips},
  {Immler}, {Soderberg}, {Sollerman}, {Boldt}, {Braithwaite}, {Brown}, {Burns},
  {Contreras}, {Covarrubias}, {Folatelli}, {Freedman}, {Gonz{\'a}lez}, {Hamuy},
  {Krzeminski}, {Madore}, {Milne}, {Morrell}, {Persson}, {Roth}, {Smith}, \&
  {Suntzeff}}]{str09}
{Stritzinger}, M., {Mazzali}, P., {Phillips}, M.~M., {et~al.} 2009, \apj, 696,
  713

\bibitem[{{Sukhbold} {et~al.}(2016){Sukhbold}, {Ertl}, {Woosley}, {Brown}, \&
  {Janka}}]{suk16}
{Sukhbold}, T., {Ertl}, T., {Woosley}, S.~E., {Brown}, J.~M., \& {Janka}, H.-T.
  2016, \apj, 821, 38

\bibitem[{{Suwa} {et~al.}(2019){Suwa}, {Tominaga}, \& {Maeda}}]{suw19}
{Suwa}, Y., {Tominaga}, N., \& {Maeda}, K. 2019, \mnras, 483, 3607

\bibitem[{{Taddia} {et~al.}(2015){Taddia}, {Sollerman}, {Leloudas},
  {Stritzinger}, {Valenti}, {Galbany}, {Kessler}, {Schneider}, \&
  {Wheeler}}]{tad15}
{Taddia}, F., {Sollerman}, J., {Leloudas}, G., {et~al.} 2015, \aap, 574, A60

\bibitem[{{Taddia} {et~al.}(2018){Taddia}, {Stritzinger}, {Bersten}, {Baron},
  {Burns}, {Contreras}, {Holmbo}, {Hsiao}, {Morrell}, {Phillips}, {Sollerman},
  \& {Suntzeff}}]{tad18a}
{Taddia}, F., {Stritzinger}, M.~D., {Bersten}, M., {et~al.} 2018, \aap, 609,
  A136

\bibitem[{{Tanaka} {et~al.}(2009){Tanaka}, {Tominaga}, {Nomoto}, {Valenti},
  {Sahu}, {Minezaki}, {Yoshii}, {Yoshida}, {Anupama}, {Benetti}, {Chincarini},
  {Della Valle}, {Mazzali}, \& {Pian}}]{tan09}
{Tanaka}, M., {Tominaga}, N., {Nomoto}, K., {et~al.} 2009, \apj, 692, 1131

\bibitem[{{Tartaglia} {et~al.}(2017){Tartaglia}, {Fraser}, {Sand}, {Valenti},
  {Smartt}, {McCully}, {Anderson}, {Arcavi}, {Elias-Rosa}, {Galbany},
  {Gal-Yam}, {Haislip}, {Hosseinzadeh}, {Howell}, {Inserra}, {Jha}, {Kankare},
  {Lundqvist}, {Maguire}, {Mattila}, {Reichart}, {Smith}, {Smith},
  {Stritzinger}, {Sullivan}, {Taddia}, \& {Tomasella}}]{tar17}
{Tartaglia}, L., {Fraser}, M., {Sand}, D.~J., {et~al.} 2017, \apj, 836, L12

\bibitem[{{Tauris} {et~al.}(2015){Tauris}, {Langer}, \&
  {Podsiadlowski}}]{tau15}
{Tauris}, T.~M., {Langer}, N., \& {Podsiadlowski}, P. 2015, \mnras, 451, 2123

\bibitem[{{Ugliano} {et~al.}(2012){Ugliano}, {Janka}, {Marek}, \&
  {Arcones}}]{ugl12}
{Ugliano}, M., {Janka}, H.-T., {Marek}, A., \& {Arcones}, A. 2012, \apj, 757,
  69

\bibitem[{{Valenti} {et~al.}(2011){Valenti}, {Fraser}, {Benetti}, {Pignata},
  {Sollerman}, {Inserra}, {Cappellaro}, {Pastorello}, {Smartt}, {Ergon},
  {Botticella}, {Brimacombe}, {Bufano}, {Crockett}, {Eder}, {Fugazza},
  {Haislip}, {Hamuy}, {Harutyunyan}, {Ivarsen}, {Kankare}, {Kotak}, {Lacluyze},
  {Magill}, {Mattila}, {Maza}, {Mazzali}, {Reichart}, {Taubenberger},
  {Turatto}, \& {Zampieri}}]{val11}
{Valenti}, S., {Fraser}, M., {Benetti}, S., {et~al.} 2011, \mnras, 416, 3138

\bibitem[{{Valenti} {et~al.}(2016)}]{val16}
{Valenti}, S. {et~al.} 2016, \mnras, 459, 3939

\bibitem[{{Van Dyk} {et~al.}(2013){Van Dyk}, {Zheng}, {Clubb}, {Filippenko},
  {Cenko}, {Smith}, {Fox}, {Kelly}, {Shivvers}, \& {Ganeshalingam}}]{van13}
{Van Dyk}, S.~D., {Zheng}, W., {Clubb}, K.~I., {et~al.} 2013, \apj, 772, L32

\bibitem[{{Van Dyk} {et~al.}(2014){Van Dyk}, {Zheng}, {Fox}, {Cenko}, {Clubb},
  {Filippenko}, {Foley}, {Miller}, {Smith}, {Kelly}, {Lee}, {Ben-Ami}, \&
  {Gal-Yam}}]{van14}
{Van Dyk}, S.~D., {Zheng}, W., {Fox}, O.~D., {et~al.} 2014, \aj, 147, 37

\bibitem[{{Wang} {et~al.}(2017){Wang}, {Cano}, {Wang}, {Zheng}, {Liu}, {Deng},
  {Yu}, {Dai}, {Han}, {Xu}, {Qiu}, {Wei}, {Li}, \& {Song}}]{Wang2017}
{Wang}, L.~J., {Cano}, Z., {Wang}, S.~Q., {et~al.} 2017, \apj, 851, 54

\bibitem[{{Woosley} {et~al.}(1973){Woosley}, {Arnett}, \& {Clayton}}]{woo73}
{Woosley}, S.~E., {Arnett}, W.~D., \& {Clayton}, D.~D. 1973, \apjs, 26, 231

\bibitem[{{Woosley} {et~al.}(1994){Woosley}, {Eastman}, {Weaver}, \&
  {Pinto}}]{woo94}
{Woosley}, S.~E., {Eastman}, R.~G., {Weaver}, T.~A., \& {Pinto}, P.~A. 1994,
  \apj, 429, 300

\bibitem[{{Yamanaka} {et~al.}(2017){Yamanaka}, {Nakaoka}, {Tanaka}, {Maeda},
  {Honda}, {Hanayama}, {Morokuma}, {Imai}, {Kinugasa}, {Murata}, {Nishimori},
  {Hashimoto}, {Gima}, {Hosoya}, {Ito}, {Karita}, {Kawabata}, {Morihana},
  {Morikawa}, {Murakami}, {Nagayama}, {Ono}, {Onozato}, {Sarugaku}, {Sato},
  {Suzuki}, {Takahashi}, {Takayama}, {Yaguchi}, {Akitaya}, {Asakura},
  {Kawabata}, {Kuroda}, {Nogami}, {Oasa}, {Omodaka}, {Saito}, {Sekiguchi},
  {Tominaga}, {Uemura}, \& {Watanabe}}]{yam17}
{Yamanaka}, M., {Nakaoka}, T., {Tanaka}, M., {et~al.} 2017, \apj, 837, 1

\bibitem[{{Yoon} {et~al.}(2019){Yoon}, {Chun}, {Tolstov}, {Blinnikov}, \&
  {Dessart}}]{yoo19}
{Yoon}, S.-C., {Chun}, W., {Tolstov}, A., {Blinnikov}, S., \& {Dessart}, L.
  2019, \apj, 872, 174

\end{thebibliography}
-----------------------------------------------

\appendix

\section{Sample properties}
\label{sec:A-1}

\begin{figure}[ht!]
 \centering
\includegraphics[width=\linewidth]{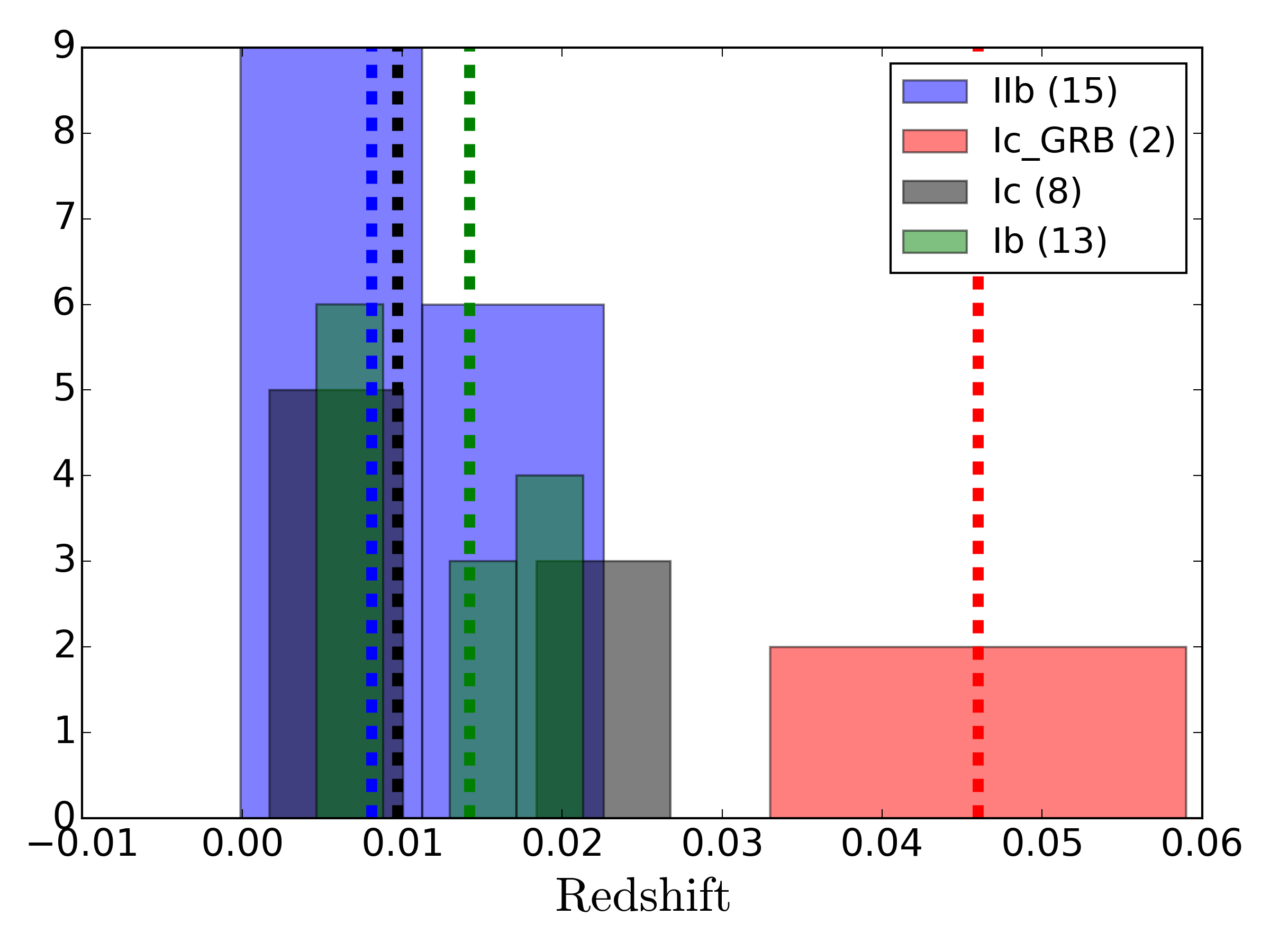}
\caption{Heliocentric redshift distribution of our $BVRIYJH$ sample. Each SN subtype is colour labeled and a vertical line of the same colour is used to mark the median of that distribution.\label{fig:z_dist}}
\end{figure}

Table~\ref{tab:table} lists the SE-SNe used in this work, together with their types, and various other relevant parameters. With the exception of a couple SNe associated with a gamma ray burst or X-ray flash, the sample has a low redshift ($z \leq 0.03$), and the redshift distribution is presented in Fig.~\ref{fig:z_dist}. The time range between the last non detection and the discovery epoch is on average less than two weeks for the sample of this work, which corresponds to an approximate mean error in explosion epochs of less than seven days. 
\begin{table*}
\centering
\scriptsize
\caption{Sample of SE~SNe used in this work. Selection criteria for this sample is described in Section 2.\label{tab:table}}
\begin{tabular}{ccccccccc}
\hline
{SN}&{Type}&{Host}&{Host redshift$^{(+)}$}&{Host $d_L^{(\dag)}$}&{MW $E(B-V)$}&{Host $E(B-V)^{(*)}$}&{$t_0$}&{References}\\
\hline
SN1993J & IIb & M81 & -0.00011 & 3.63 & 0.07 & 0.10 & 49073.50 & (a) \\
SN2004ex & IIb & NGC0182 & 0.01755 & 70.60 & 0.02 & 0.08 & 53287.90 & (CSP) \\
SN2004ff & IIb & ESO-552-G040 & 0.0226 & 92.70 & 0.03 & 0.10 & 53297.66 &(CSP)\\
SN2004gq & Ib & NGC1832 & 0.006468 & 25.10 & 0.06 & 0.08 & 53346.87 & (CSP),(Cfa)\\
SN2004gv & Ib & NGC0856 & 0.019973 & 79.60 & 0.03 & 0.03 & 53345.27 & (CSP) \\
SN2005aw & Ic & IC4837A & 0.009498 & 41.50 & 0.05 & 0.21 & 53445.67 & (CSP)\\
SN2005em & Ic & IC0307 & 0.025981 & 105.00 & 0.08 & 0.00 & 53635.00 & (CSP)\\
SN2005hg & Ib & UGC1394 & 0.02131 & 86.00 & 0.09 & None & 53665.75 & (CSP)\\
SN2005kl & Ic & NGC4369 & 0.003485 & 21.57 & 0.02 & None & 53686.14 & (Cfa)\\
SN2005mf & Ic & UGC4798 & 0.02676 & 113.00 & 0.01 & None & 53723.33 & (Cfa)\\
SN2006aj & Ic-GRB & A032139+1652 & 0.033 & 132.40 & 0.13 & 0.00 & 53784.15 & (Cfa),(b)\\
SN2006ba & IIb & NGC2980 & 0.01908 & 82.70 & 0.04 & 0.10 & 53801.11 & (Cfa),(CSP)\\
SN2006ep & Ib & NGC0214 & 0.015134 & 61.90 & 0.03 & None & 53975.49 & (Cfa),(CSP)\\
SN2006fo & Ib & UGC02019 & 0.020698 & 82.70 & 0.02 & 0.21 & 53983.36 & (Cfa),(CSP)\\
SN2006lc & Ib & NGC7364 & 0.016228 & 59.20 & 0.06 & 0.36 & 54014.74 & (Cfa),(CSP)\\
SN2006T & IIb & NGC3054 & 0.008091 & 31.60 & 0.06 & 0.14 & 53757.64 & (Cfa),(CSP)\\
SN2007C & Ib & NGC4981 & 0.005604 & 21.00 & 0.04 & 0.43 & 54095.44 & (Cfa),(CSP)\\
SN2007gr & Ic & NGC1058 & 0.0017 & 9.29 & 0.09 & 0.03 & 54325.00 & (Cfa),(c)\\
SN2007kj & Ib & NGC7803 & 0.017899 & 72.50 & 0.07 & 0.00 & 54363.61 & (Cfa),(CSP)\\
SN2007uy & Ib & NGC2770 & 0.0065 & 31.33 & 0.02 & 0.63 & 54462.33 & (Cfa),(d)\\
SN2007Y & Ib & NGC1187 & 0.004637 & 18.40 & 0.02 & 0.00 & 54145.00 & (CSP),(e)\\
SN2008aq & IIb & MCG-02-33-020 & 0.007972 & 26.90 & 0.04 & 0.00 & 54510.79 & (Cfa),(CSP)\\
SN2008ax & IIb & NGC4490 & 0.0019 & 9.20 & 0.02 & 0.28 & 54528.30 & (Cfa),(f)\\
SN2008D & Ib & NGC2770 & 0.0065 & 31.33 & 0.02 & 0.63 & 54474.50 & (Cfa),(g)\\
SN2008hh & Ic & IC0112 & 0.01941 & 77.70 & 0.04 & 0.00 & 54780.69 & (Cfa),(CSP)\\
SN2009bb & Ic-BL & NGC3278 & 0.00988 & 39.80 & 0.09 & 0.48 & 54909.10 & (CSP),(h)\\
SN2009iz & Ib & UGC02175 & 0.01419 & 58.60 & 0.07 & None & 55083.00 & (Cfa)\\
SN2009jf & Ib & NGC7479 & 0.0079 & 33.73 & 0.10 & 0.05 & 55099.00 & (Cfa),(i)\\
SN2009K & IIb & NGC1620 & 0.011715 & 44.10 & 0.05 & 0.06 & 54843.57 & (Cfa),(CSP)\\
SN2010as & IIb & NGC6000 & 0.0073 & 27.16 & 0.15 & 0.42 & 55270.75 & (j)\\
SN2010bh & Ic-GRB & A071031-5615 & 0.059 & 244.34 & 0.10 & 0.14 & 55271.53 & (k)\\
SN2011fu & IIb & UGC1626 & 0.019 & 74.47 & 0.07 & 0.01 & 55824.00 & (l)\\
SN2011dh & IIb & M51 & 0.002 & 7.87 & 0.03 & 0.05 & 55712.50 & (m)\\
SN2011hs & IIb & IC5267 & 0.0057 & 24.10 & 0.01 & 0.16 & 55871.50 & (n)\\
SN2013df & IIb & NGC4414 & 0.0024 & 21.37 & 0.02 & 0.08 & 56447.30 & (o)\\
SN2016coi & Ic-BL & UGC11868 & 0.00364 & 17.20 & 0.07 & 0.12 & 57532.50 & (p)\\
SN2017czd & IIb & UGC9567 & 0.00835 & 32.00 & 0.02 & 0.00 & 57845.00 & (q)\\
\hline
\end{tabular}
\tablefoot{$^{(+)}$: Heliocentric redshift. $^{(\dag)}$: Luminosity distance, in Mpc. $^{(*)}:$ If no published value is found, we quote ``None''. Zero values are consistent with no host reddening, as published in the proper references. (CSP): \cite{tad18a}, (Cfa): \cite{bia14}, (a): \cite{ric94}, (b): \cite{mir06}, (c): \cite{hun09}, (d): \cite{roy13}, (e): \cite{str09}, (f): \cite{pas08}, (g): \cite{maz08,tan09} (h): \cite{pig11}, (i): \cite{val11}, (j): \cite{fol14}, (k): \cite{can11,oli12} , (l): \cite{mor15}, (m): \cite{arc11,sah13,erg15}, (n): \cite{buf14}, (o): \cite{van14}, (p): \cite{yam17,pre18b}, (q): \cite{nak19}}

\end{table*}

\begin{table*}
\centering
\scriptsize
\caption{Peak parameters of our $BVRIYJH$ light curves and the $^{56}$Ni masses obtained as described in Section \ref{sec:analysis}. All our luminosities and nickel masses are lower limits, as described in the manuscript. For the \cite{Khatami19} $^{56}$Ni values we use their reccomended $\beta$ values, of 0.82 for SNe~IIb, and $9/8$ for SNe~Ib and SNe~Ic (including SNe~Ic-BL).\label{tab:table2}}
\begin{tabular}{cccccccc}
\hline
{SN}&{Type}& {$t_p \ [days]$} & {$L_p \ [10^{41} erg/s]$} & {Arnett \ $[M_{\odot}]$} & {K\&K \ $[M_{\odot}]$} & {Tail \ $[M_{\odot}]$} \\
\hline
SN1993J & IIb & 21.89 & 17.20 & 0.10 & 0.06 & 0.05\\
SN2004ex & IIb & 20.86 & 13.34 & 0.08 & 0.04 & 0.05\\
SN2004ff & IIb & 15.33 & 19.00 & 0.08 & 0.05 & None\\
SN2004gq & Ib & 12.42 & 13.81 & 0.05 & 0.04 & 0.05\\
SN2004gv & Ib & 22.38 & 18.79 & 0.12 & 0.08 & None\\
SN2005aw & Ic & 12.01 & 27.48 & 0.10 & 0.07 & None\\
SN2005em & Ic & 13.43 & 15.66 & 0.06 & 0.05 & None\\
SN2005hg & Ib & 18.93 & 19.63 & 0.10 & 0.07 & 0.07\\
SN2005kl & Ic & 18.49 & 6.57 & 0.03 & 0.02 & 0.02\\
SN2005mf & Ic & 10.98 & 15.45 & 0.05 & 0.04 & None\\
SN2006aj & Ic-GRB & 10.18 & 50.24 & 0.15 & 0.12 & None\\
SN2006ba & IIb & 20.91 & 13.77 & 0.08 & 0.04 & None\\
SN2006ep & Ib & 14.58 & 10.32 & 0.04 & 0.03 & None\\
SN2006fo & Ib & 24.53 & 28.93 & 0.20 & 0.13 & None\\
SN2006lc & Ib & 25.98 & 17.11 & 0.12 & 0.08 & None\\
SN2006T & IIb & 23.53 & 16.19 & 0.11 & 0.06 & 0.05\\
SN2007C & Ib & 20.45 & 10.94 & 0.06 & 0.04 & 0.03\\
SN2007gr & Ic & 13.22 & 12.97 & 0.05 & 0.04 & 0.03\\
SN2007kj & Ib & 18.16 & 10.35 & 0.05 & 0.04 & 0.03\\
SN2007uy & Ib & 19.36 & 37.94 & 0.21 & 0.14 & 0.10\\
SN2007Y & Ib & 20.18 & 4.81 & 0.03 & 0.02 & None\\
SN2008aq & IIb & 20.59 & 6.25 & 0.04 & 0.02 & 0.02\\
SN2008ax & IIb & 21.87 & 7.68 & 0.05 & 0.02 & 0.02\\
SN2008D & Ib & 20.42 & 13.96 & 0.08 & 0.05 & 0.04\\
SN2008hh & Ic & 11.18 & 17.38 & 0.06 & 0.04 & None\\
SN2009bb & Ic-BL & 12.77 & 44.62 & 0.17 & 0.12 & 0.08\\
SN2009iz & Ib & 28.84 & 12.85 & 0.10 & 0.07 & 0.07\\
SN2009jf & Ib & 21.65 & 28.16 & 0.17 & 0.12 & 0.11\\
SN2009K & IIb & 27.06 & 19.95 & 0.15 & 0.08 & None\\
SN2010as & IIb & 17.68 & 30.46 & 0.15 & 0.09 & 0.09\\
SN2010bh & Ic-GRB & 9.24 & 37.86 & 0.11 & 0.09 & None\\
SN2011fu & IIb & 22.30 & 23.24 & 0.14 & 0.08 & None\\
SN2011dh & IIb & 20.25 & 11.01 & 0.06 & 0.03 & 0.04\\
SN2011hs & IIb & 17.29 & 8.49 & 0.04 & 0.02 & None\\
SN2013df & IIb & 21.92 & 16.28 & 0.10 & 0.05 & 0.06\\
SN2016coi & Ic-BL & 19.49 & 25.87 & 0.14 & 0.10 & 0.10\\
SN2017czd & IIb & 14.50 & 6.22 & 0.03 & 0.02 & None\\

\hline 
\end{tabular}
\end{table*}

\subsection{SE~SN rise times and $^{56}$Ni systematics as measured at the peak.}

\begin{figure}[ht!]
 \centering
\includegraphics[width=\linewidth]{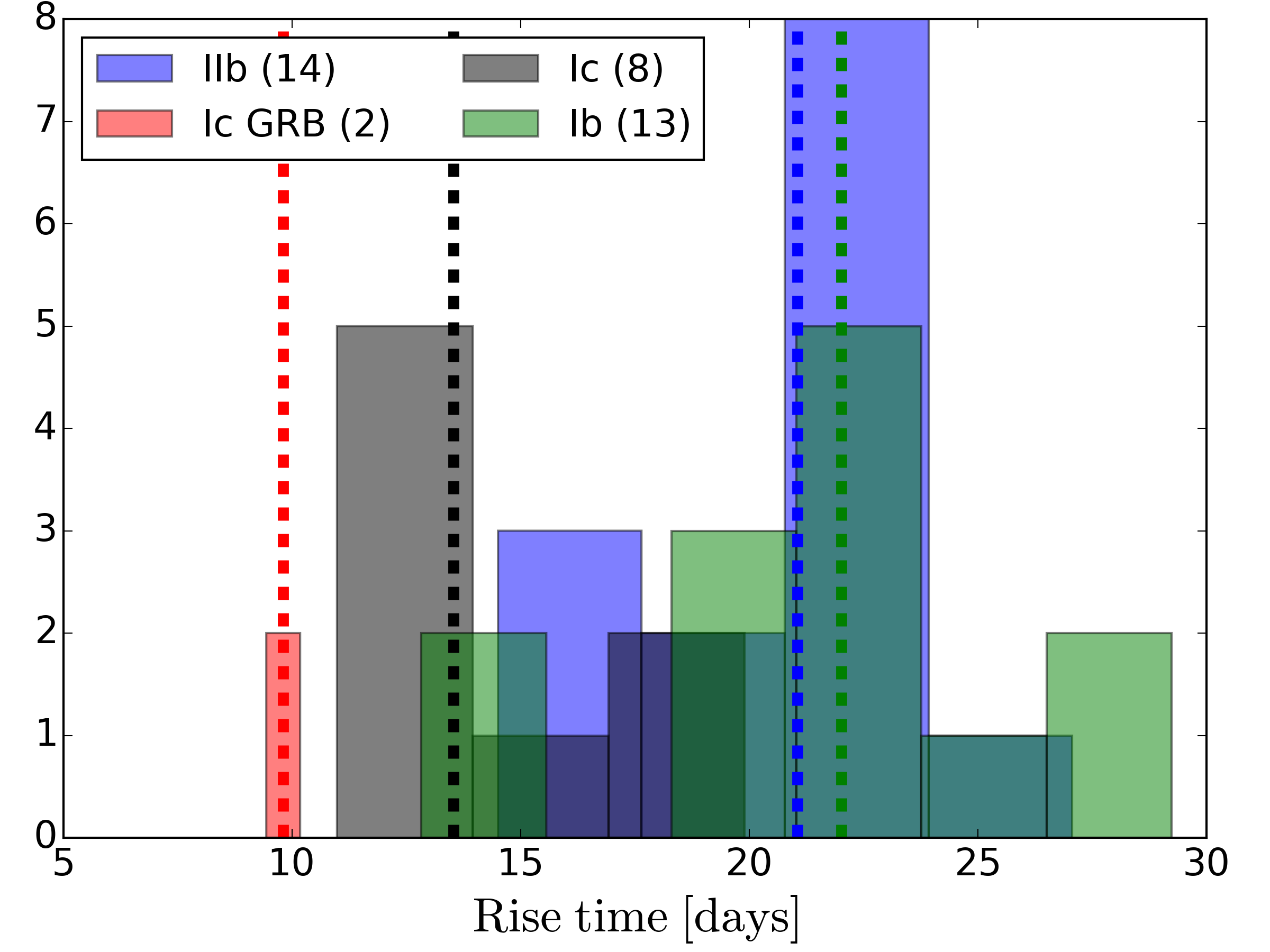}
\caption{Rise time distribution of our $BVRIYJH$ sample. Each SN subtype is colour labeled and a vertical line of the same colour is used to mark the median of that distribution.\label{fig:rise_dist}}
\end{figure}

\begin{figure}[ht!]
 \centering
\includegraphics[width=\linewidth]{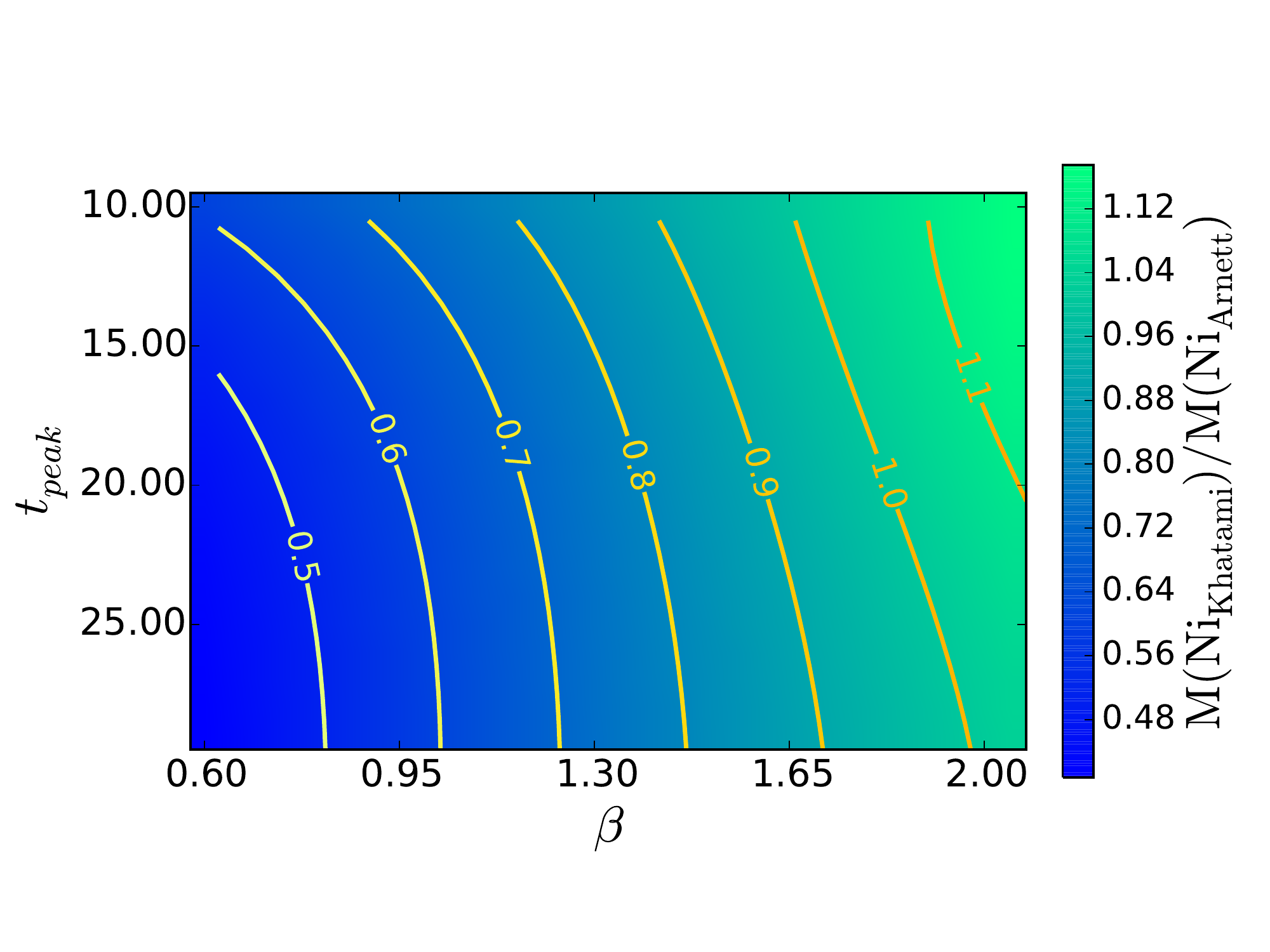}
\caption{Colour plot showing the ratio of the $^{56}$Ni mass measured using the \cite{Khatami19} method to the one measured with the Arnett's rule, as a function of the peak time and the $\beta$ parameter.\label{fig:khatami_arnett}}
\end{figure}

In Fig.~\ref{fig:rise_dist} we show the rise time distributions obtained for the different SNe sub-types. As expected, the shorter rise times belong to the SNe~Ic (including Ic-BL) while the rise times for SN types Ib/IIb are higher.
The effect of the rise time and mixing (as measured by the $\beta$ parameter) on the ratio of the $^{56}$Ni mass measured with the Khatami \& Kasen and the Arnett method is shown in Fig.~\ref{fig:khatami_arnett}, where we show a 2D colour plot with the dependence of this ratio. 
In most of the parameter space Arnett's rule gives a relative overestimate (compared to \citeauthor{Khatami19}), while for short rise times and higher mixing Arnett's rule is comparable to the Khatami \& Kasen method. Considering the rise times of our sample, we expect that the parameter space where Arnett's rule is more accurate is covered by SN types Ic/Ic-BL SNe, while for SNe~IIb we expect that Arnett's rule will always be an overestimation.

\newpage

\section{Nebular bolometric corrections for SNe~II}
\label{sec:B}
\cite{ber09} explored bolometric corrections (BCs) at nebular epochs for SN~II, finding that considering the $U-K$ bands the BC of SN~1987A roughly agrees with the SNe~II 1999em and 2003hn, within a range of $\pm0.25$ mags. We further test this conclusion here.\\ 
\indent We used nebular-phase spectra of SN~1987A and a sample of 18 SNe~II, including one peculiar object (SN~2009E). Together with nebular spectra from  models of \cite{jer12} and \cite{des13}, we cover a wide range in expected physical progenitor properties, such as progenitor mass and metalicity, for red supergiants (the presumed progenitors of SNe~II). Spectra used cover from 150 to 300 days past explosion with a wavelength range of 3500-9000\AA. With this sample the fraction of integrated flux in the range 3500-9000\AA\ ($F(3500-9000)$) with respect to the flux in the $V$ band was estimated, which we name the pBC (pseudo-bolometric-correction):
\begin{eqnarray}
    pBC = -2.5 log(F(3500-9000)/F_V)
\end{eqnarray}
The estimated pBCs from observations span $\pm0.15$ magnitudes around the correction for SN~1987A (Fig.~\ref{fig:BCs}.1), which is very close to the sample mean. The pBC mean of the models is 0.12 mags from SN1987A and observations. We found that the $3\sigma$ dispersion from observations is 0.26 mags, which translates to a dispersion of $\approx 25\%$ in SN~II $^{56}$Ni mass. Although significant, this cannot explain the differences obtained in this work, because the deviations are not systematically skewed toward the direction that would make SN~II $^{56}$Ni masses larger. Thus, through this analysis we conclude that using the bolometric correction from SN~1987A to estimate SN~II $^{56}$Ni masses does not produce a systematic that is biasing our results.

\begin{figure}[ht!]
 \centering
\includegraphics[width=\linewidth]{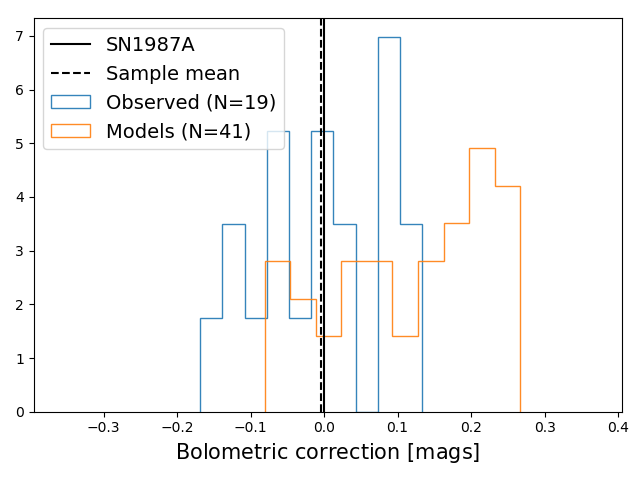}
\caption{Pseudo-bolometric corrections at the nebular phase for SNe~II. These were calculated using observed spectra from 18 SNe, with epochs ranging from 150 to 300 days past explosion and models from \citep{jer12} and \cite{des13}.}
\end{figure}
\label{fig:BCs}
\newpage

\section{$BVRIYJH$ pseudo-bolometric light curves and $^{56}$Ni mass measurements}
\label{apB}
We now present the full sample of pseudo-bolometric light curves of our SE-SNe, obtained from the integration of the flux in the $BVRIYJH$ (or equivalent) bands as described in Section \ref{sec:analysis}. These light curves are presented in Figs B.1 through B.7.

\vspace{20cm}
\clearpage
\newpage

\begin{figure*}[ht!]
\centering
\includegraphics[width=9cm]{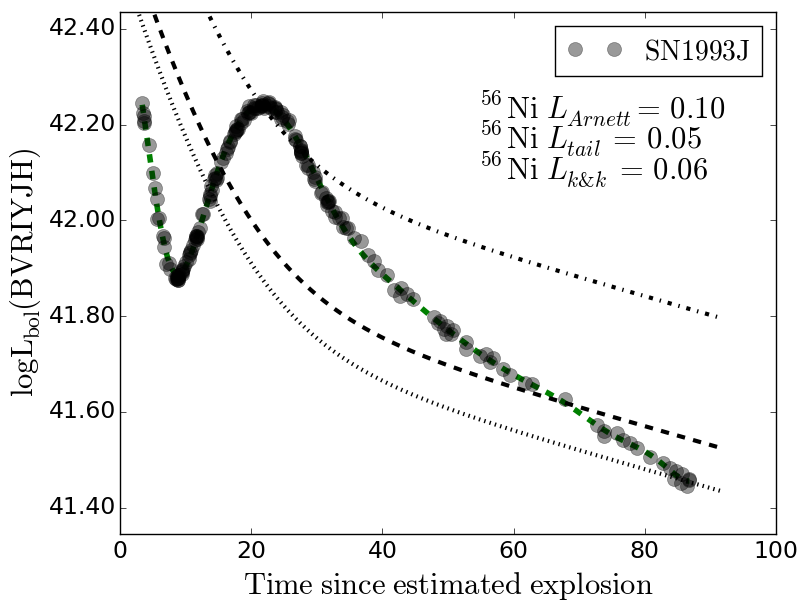}
\includegraphics[width=9cm]{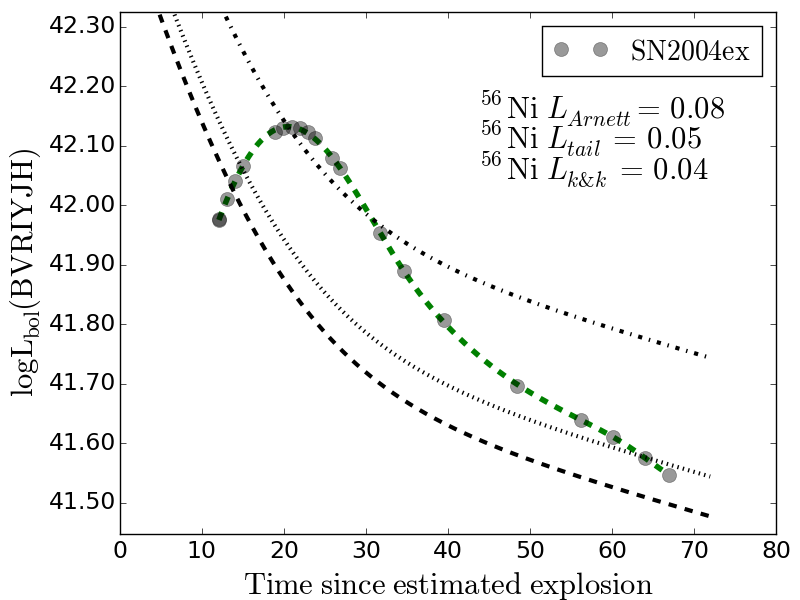}
\includegraphics[width=9cm]{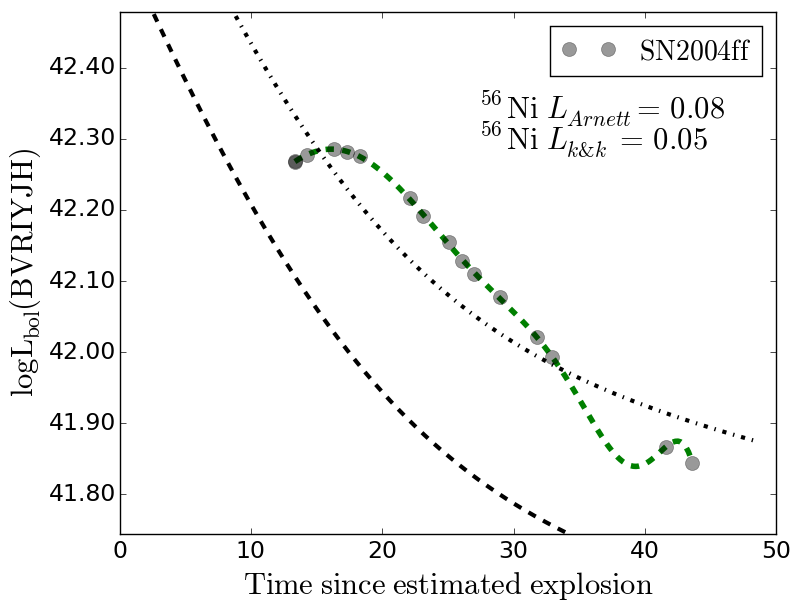}
\includegraphics[width=9cm]{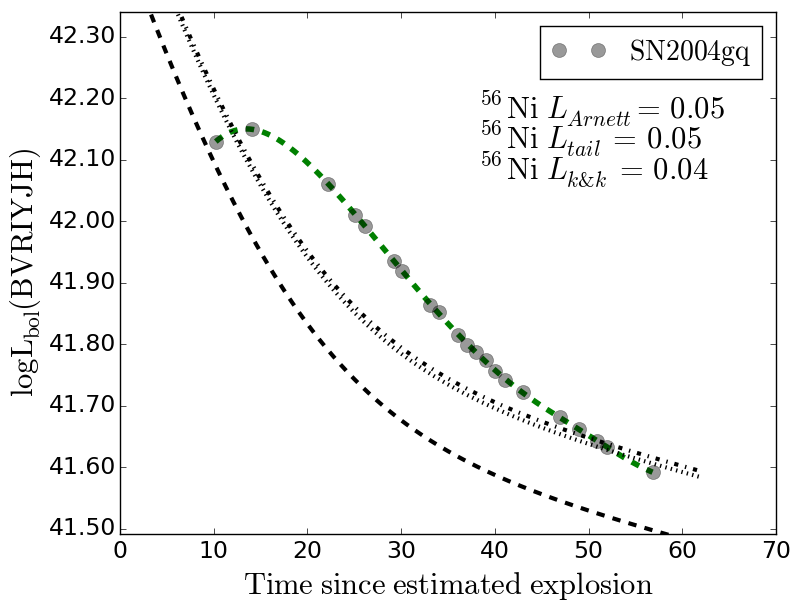}
\includegraphics[width=9cm]{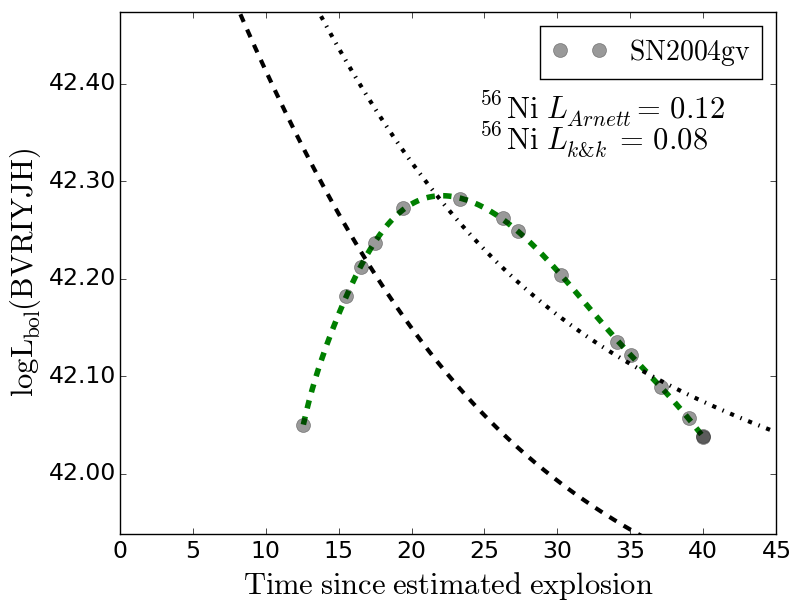}
\includegraphics[width=9cm]{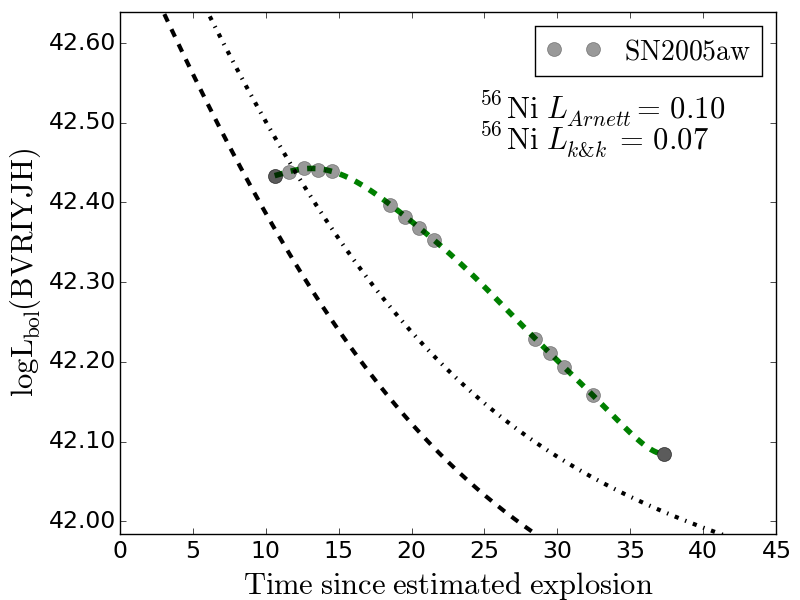}
\caption{$BVRIYJH$ pseudo-bolometric light curves for SE-SNe. The dotted lines give the $^{56}$Ni mass decay curve for that estimated through the Tail. The dashed line gives the $^{56}$Ni mass decay curve from \citeauthor{Khatami19}, while the dot-dashed line gives that from Arnett.}
\end{figure*}

\begin{figure*}[ht!]
\centering
\includegraphics[width=9cm]{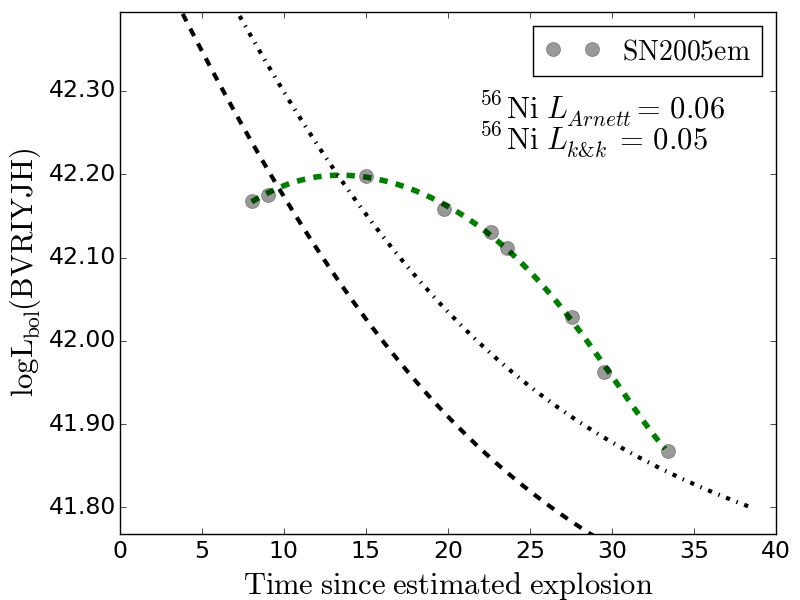}
\includegraphics[width=9cm]{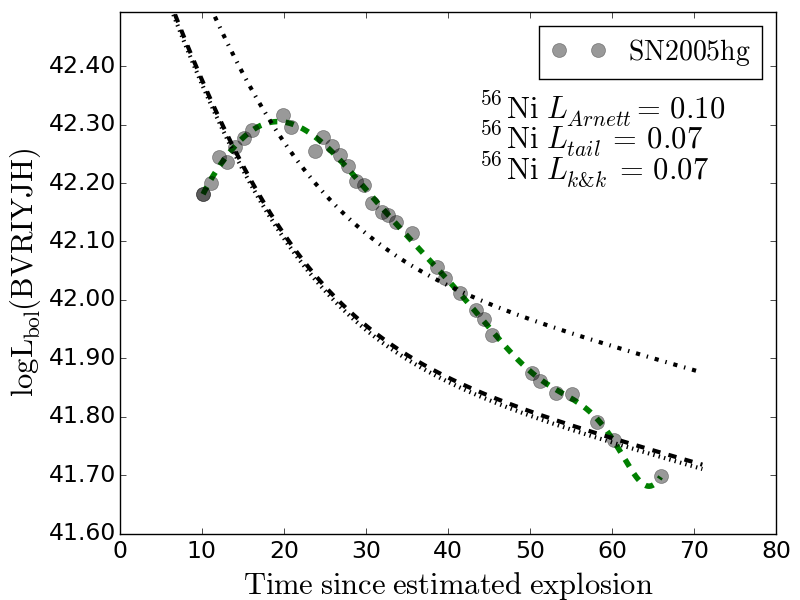}
\includegraphics[width=9cm]{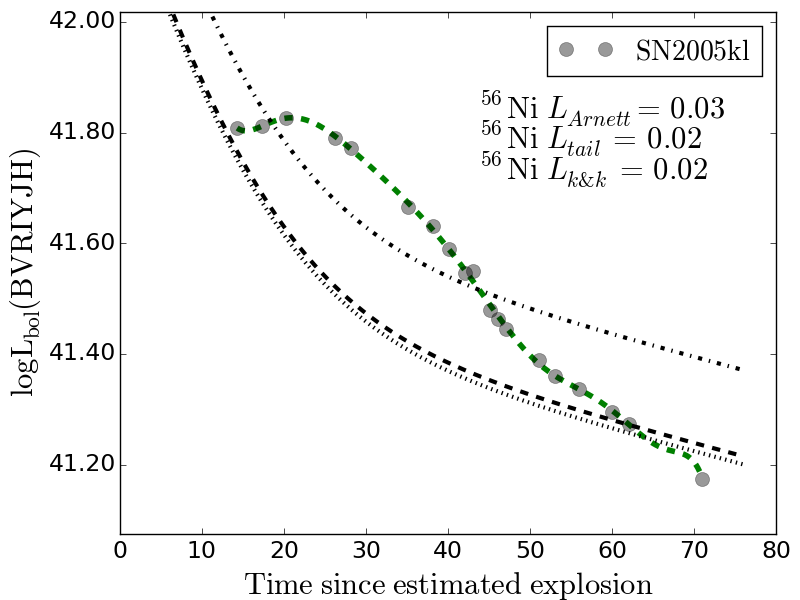}
\includegraphics[width=9cm]{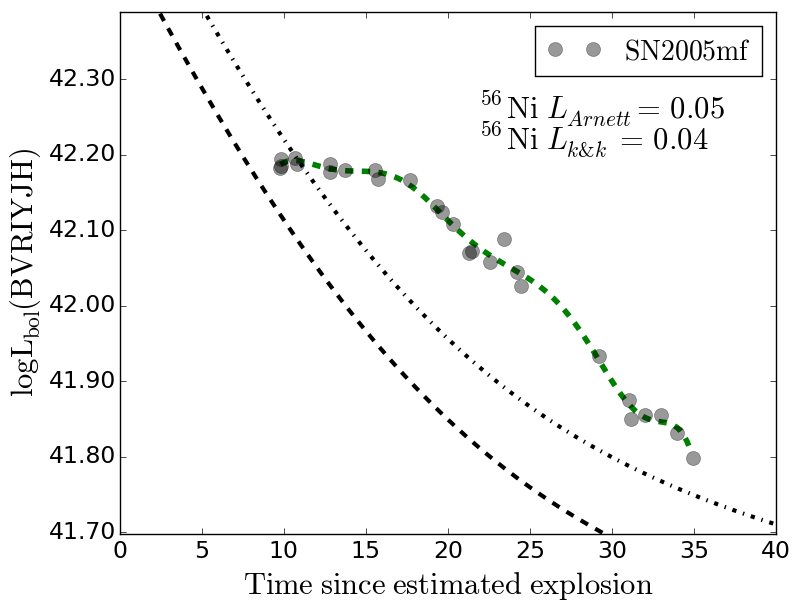}
\includegraphics[width=9cm]{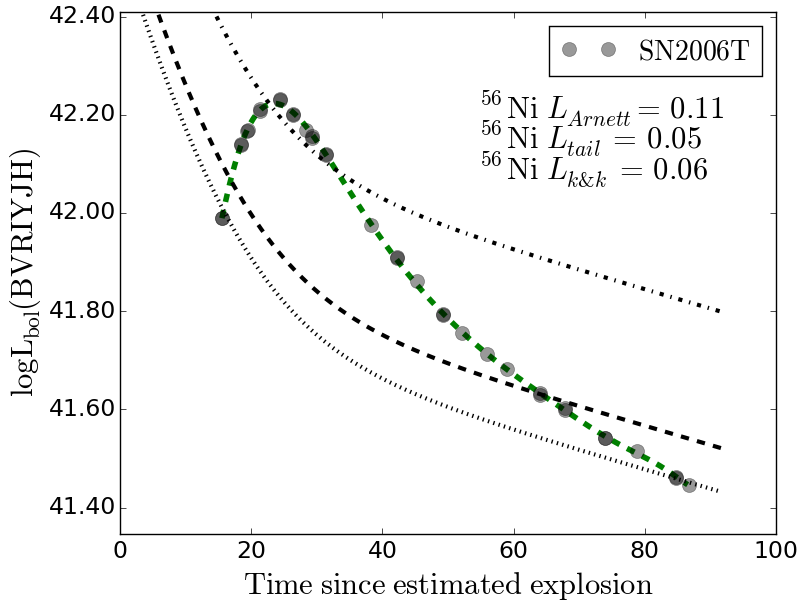}
\includegraphics[width=9cm]{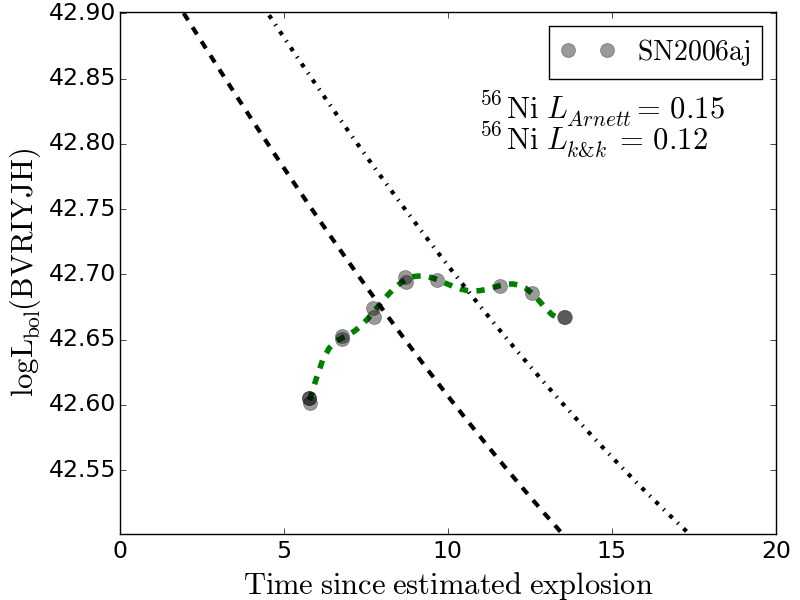}
\caption{$BVRIYJH$ pseudo-bolometric light curves for SE-SNe. The dotted lines give the $^{56}$Ni mass decay curve for that estimated through the Tail. The dashed line gives the $^{56}$Ni mass decay curve from \citeauthor{Khatami19}, while the dot-dashed line gives that from Arnett.}
\end{figure*}

\begin{figure*}[ht!]
 \centering
\includegraphics[width=9cm]{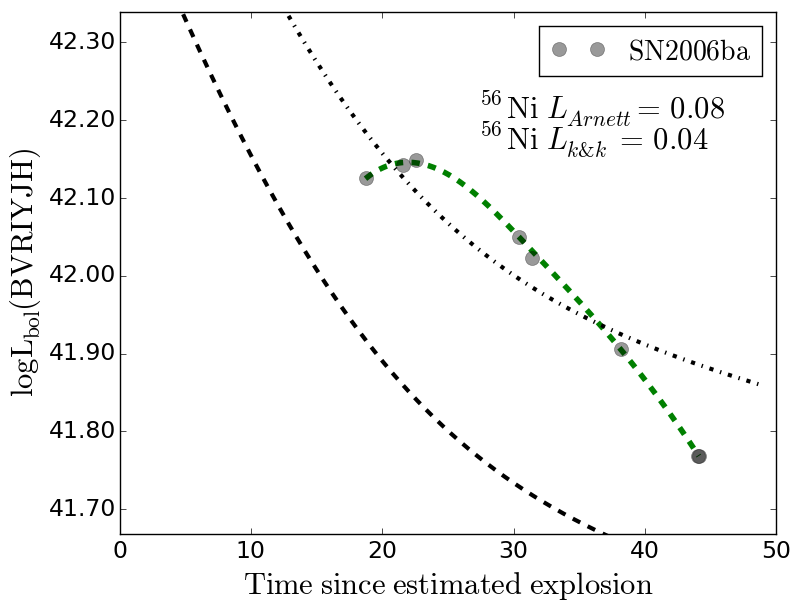}
\includegraphics[width=9cm]{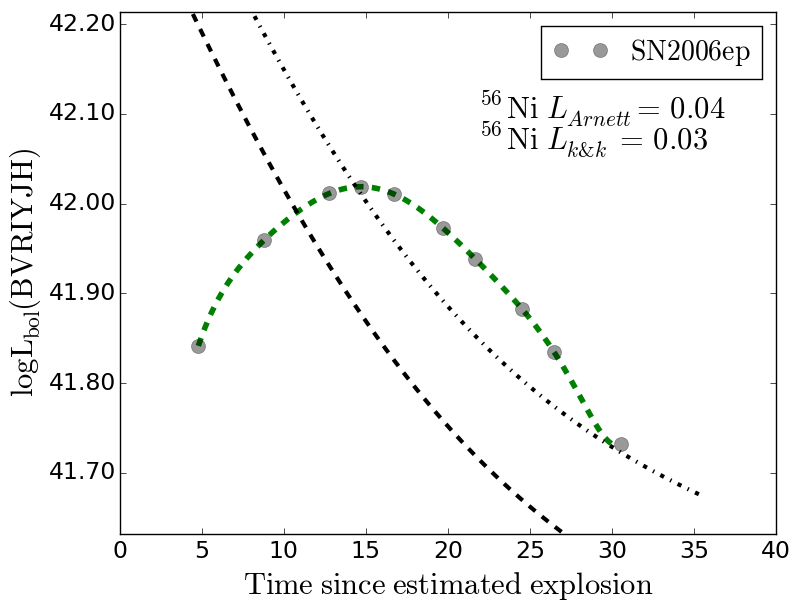}
\includegraphics[width=9cm]{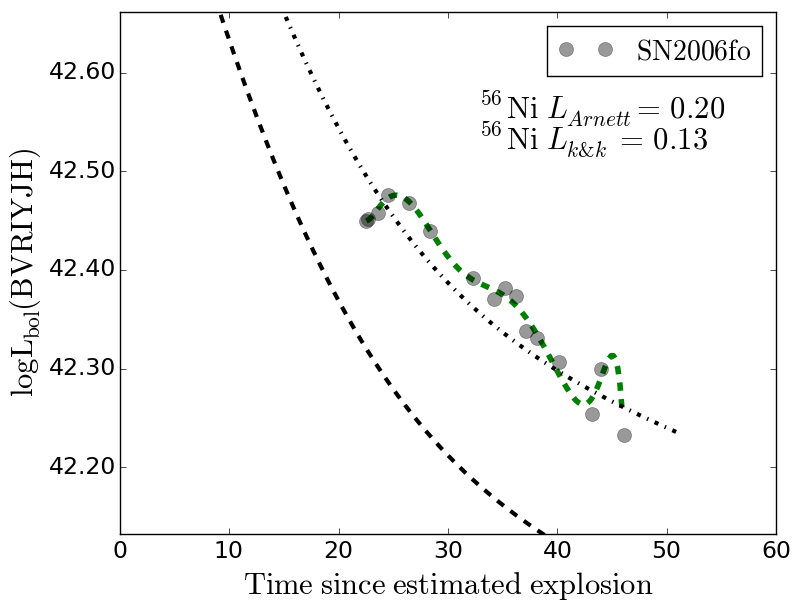}
\includegraphics[width=9cm]{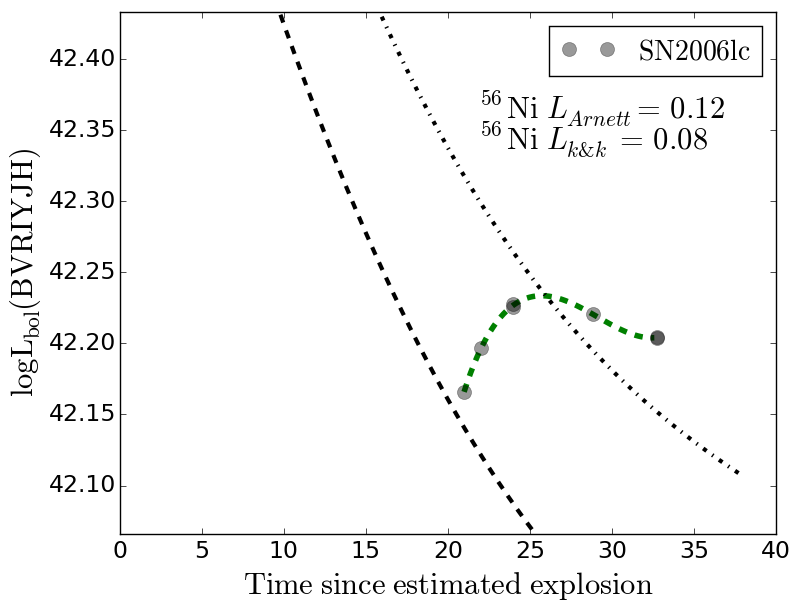}
\includegraphics[width=9cm]{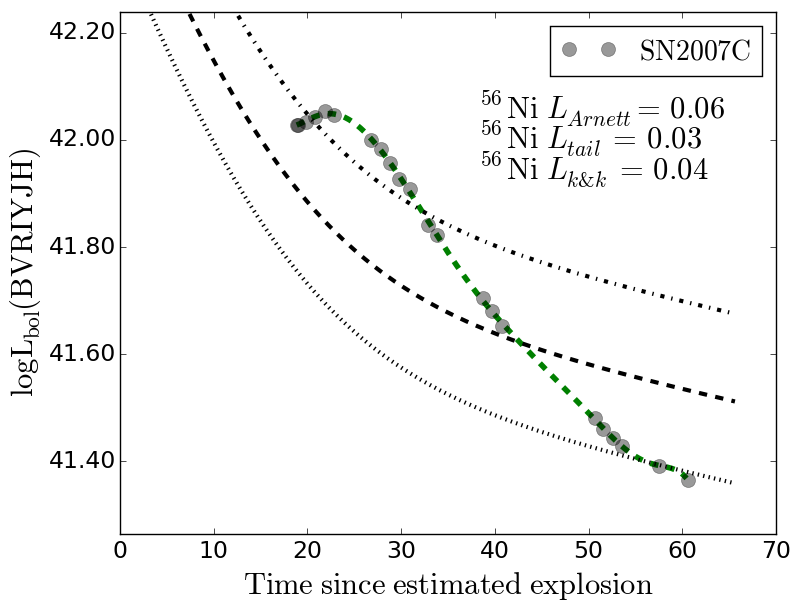}
\includegraphics[width=9cm]{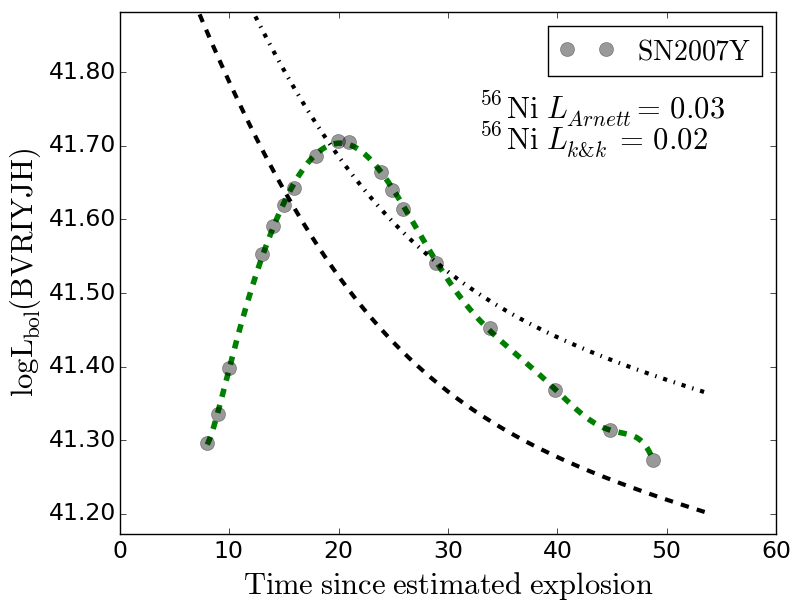}
\caption{$BVRIYJH$ pseudo-bolometric light curves for SE-SNe. The dotted lines give the $^{56}$Ni mass decay curve for that estimated through the Tail. The dashed line gives the $^{56}$Ni mass decay curve from \citeauthor{Khatami19}, while the dot-dashed line gives that from Arnett.}
\end{figure*}

\begin{figure*}[ht!]
 \centering
\includegraphics[width=9cm]{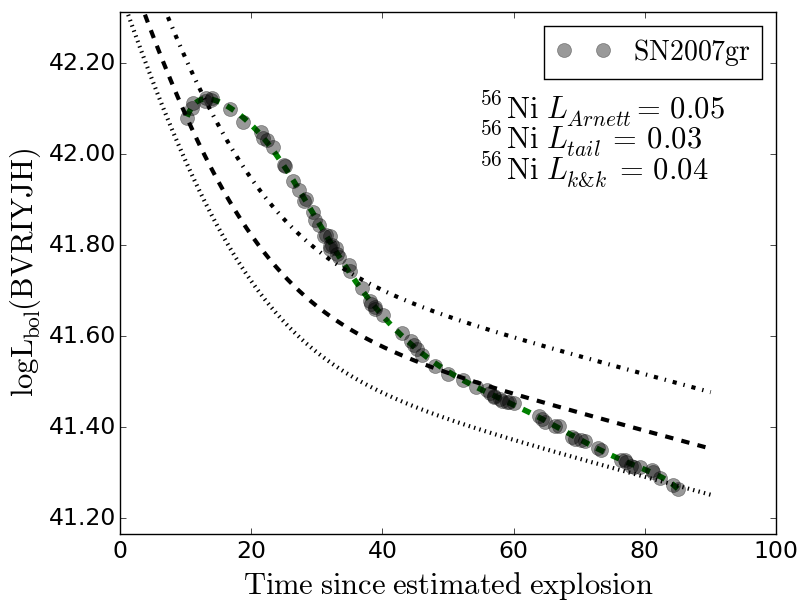}
\includegraphics[width=9cm]{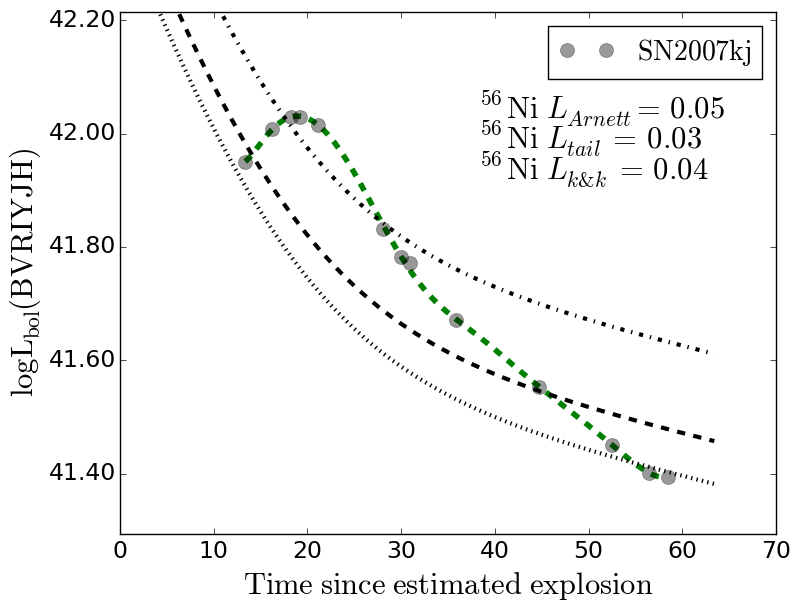}
\includegraphics[width=9cm]{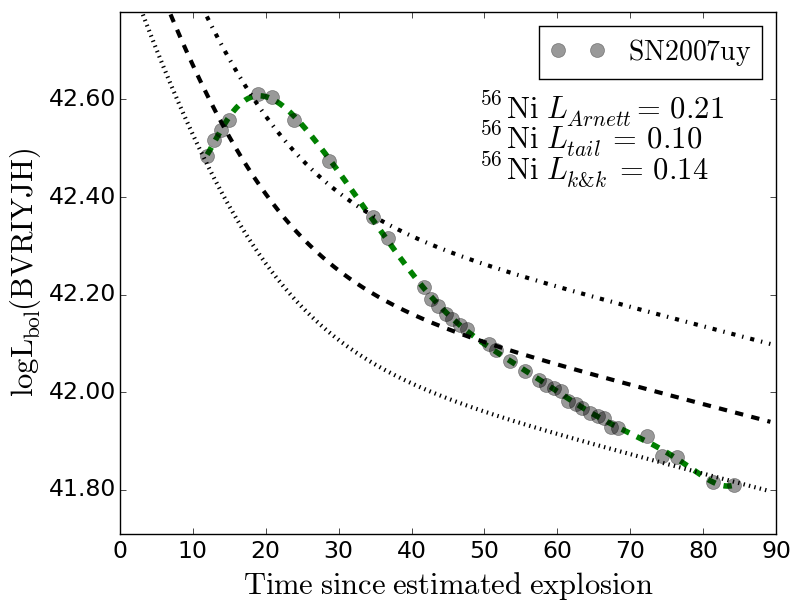}
\includegraphics[width=9cm]{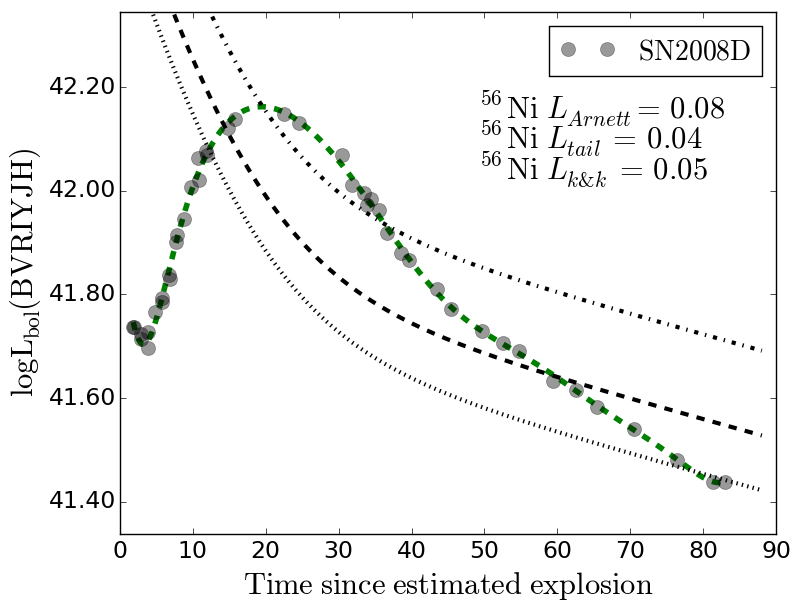}
\includegraphics[width=9cm]{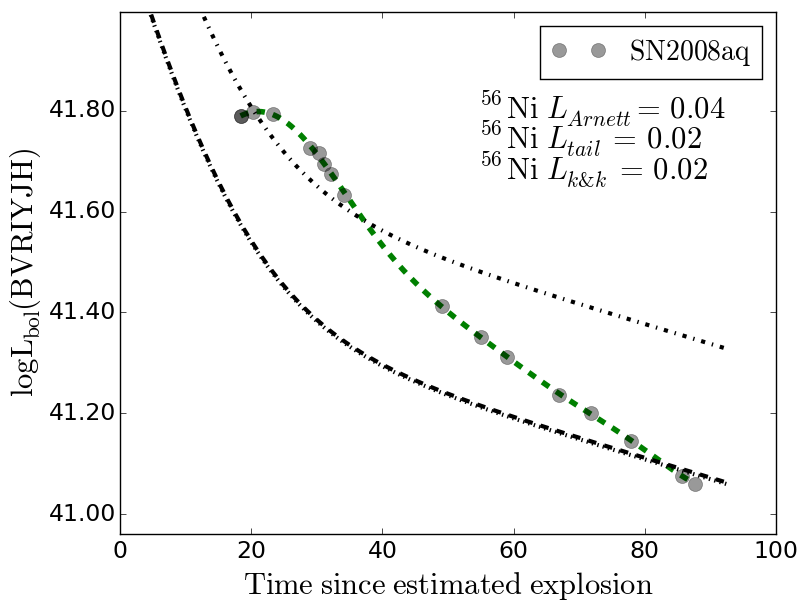}
\includegraphics[width=9cm]{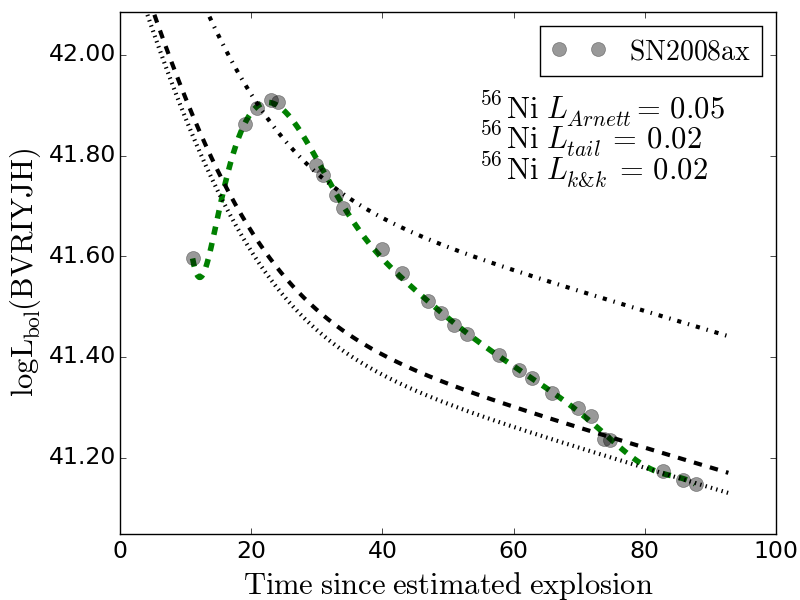}
\caption{$BVRIYJH$ pseudo-bolometric light curves for SE-SNe. The dotted lines give the $^{56}$Ni mass decay curve for that estimated through the Tail. The dashed line gives the $^{56}$Ni mass decay curve from \citeauthor{Khatami19}, while the dot-dashed line gives that from Arnett.}
\end{figure*}

\begin{figure*}[ht!]
 \centering
\includegraphics[width=9cm]{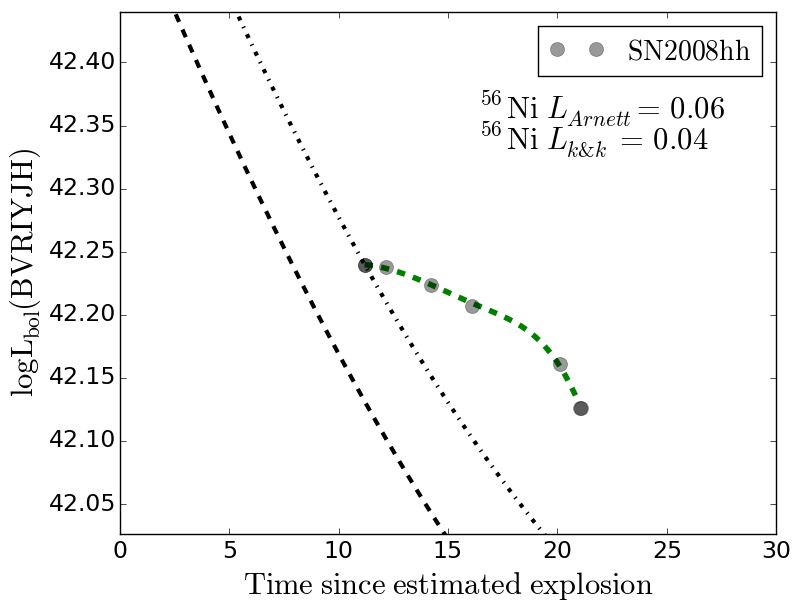}
\includegraphics[width=9cm]{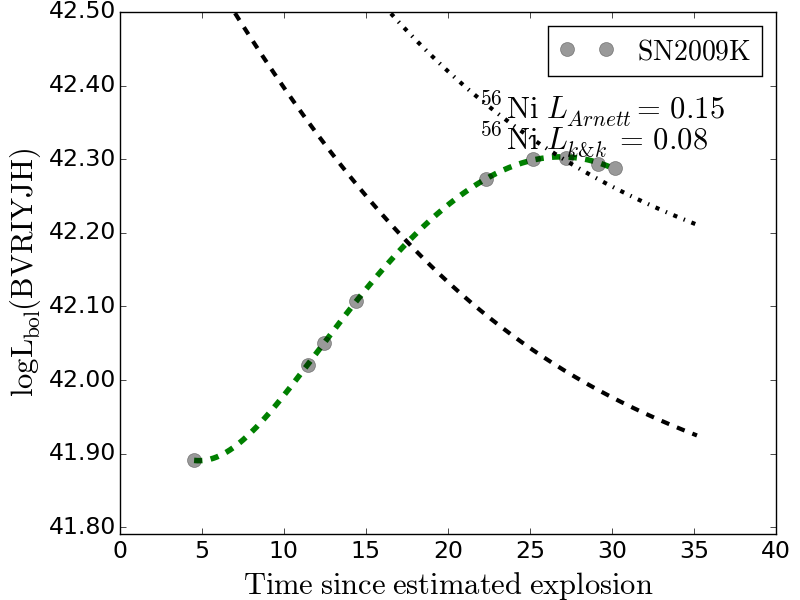}
\includegraphics[width=9cm]{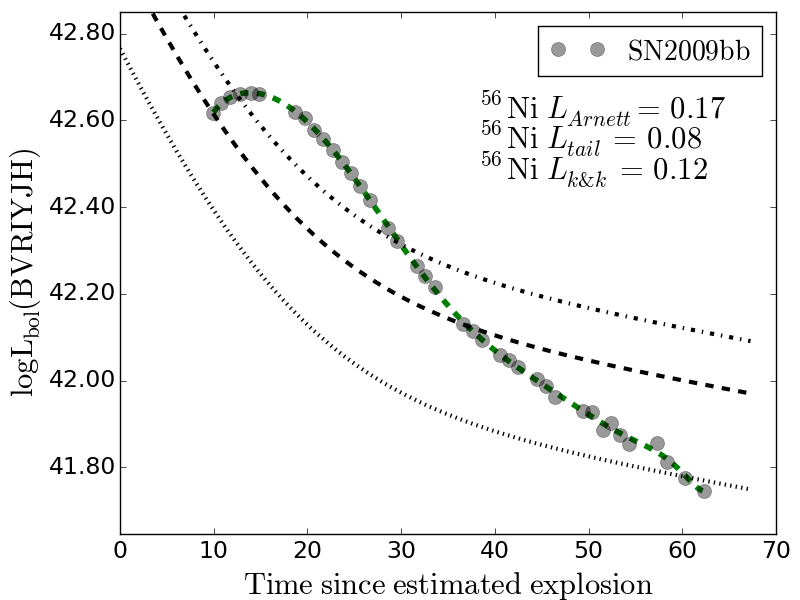}
\includegraphics[width=9cm]{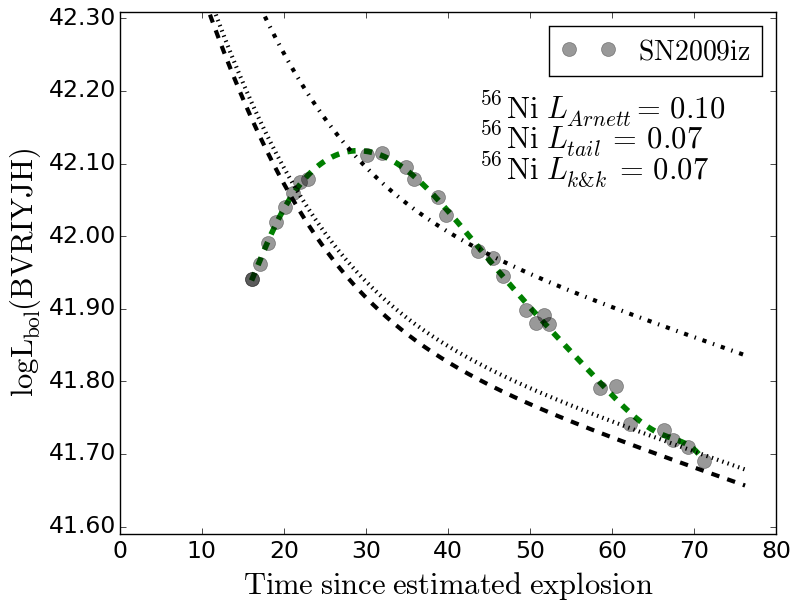}
\includegraphics[width=9cm]{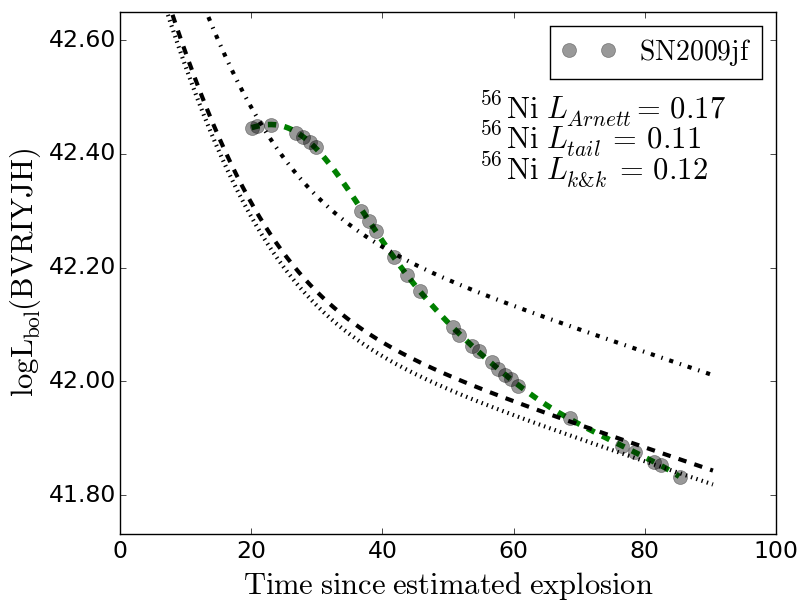}
\includegraphics[width=9cm]{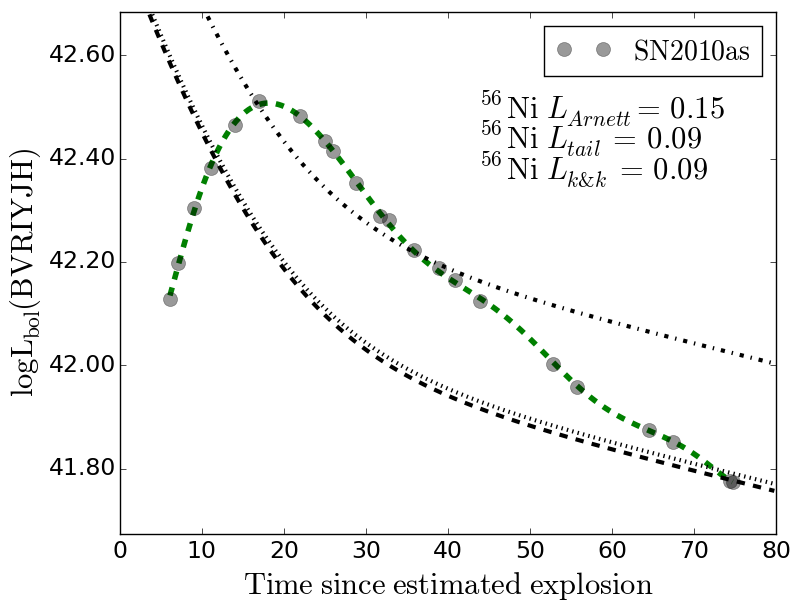}
\caption{$BVRIYJH$ pseudo-bolometric light curves for SE-SNe. The dotted lines give the $^{56}$Ni mass decay curve for that estimated through the Tail. The dashed line gives the $^{56}$Ni mass decay curve from \citeauthor{Khatami19}, while the dot-dashed line gives that from Arnett.}
\end{figure*}

\begin{figure*}[ht!]
 \centering
\includegraphics[width=9cm]{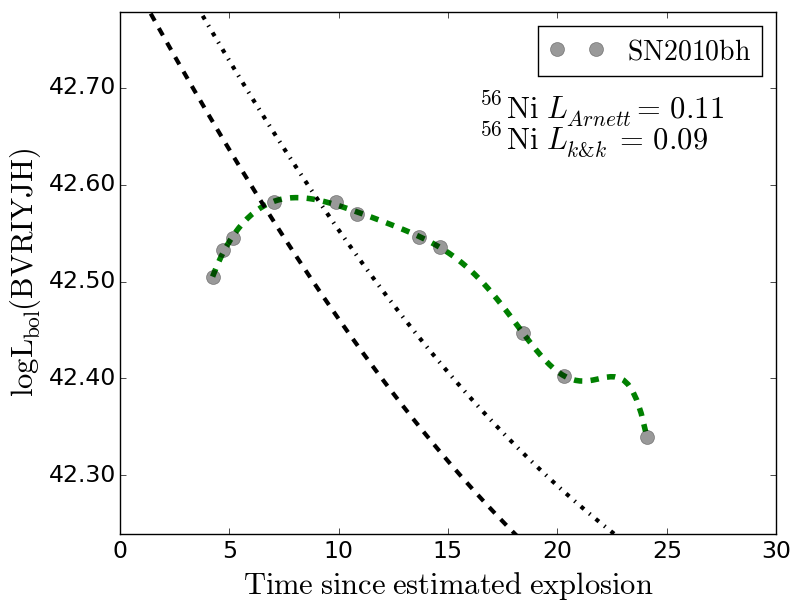}
\includegraphics[width=9cm]{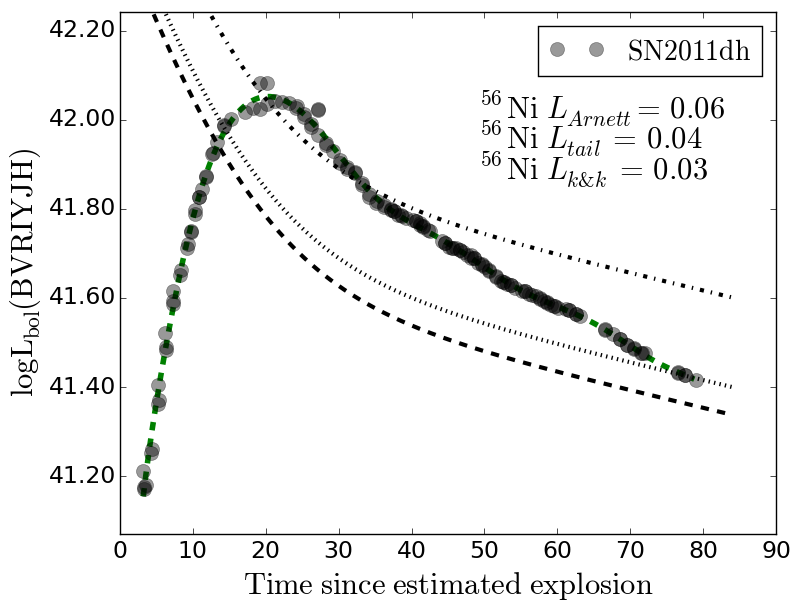}
\includegraphics[width=9cm]{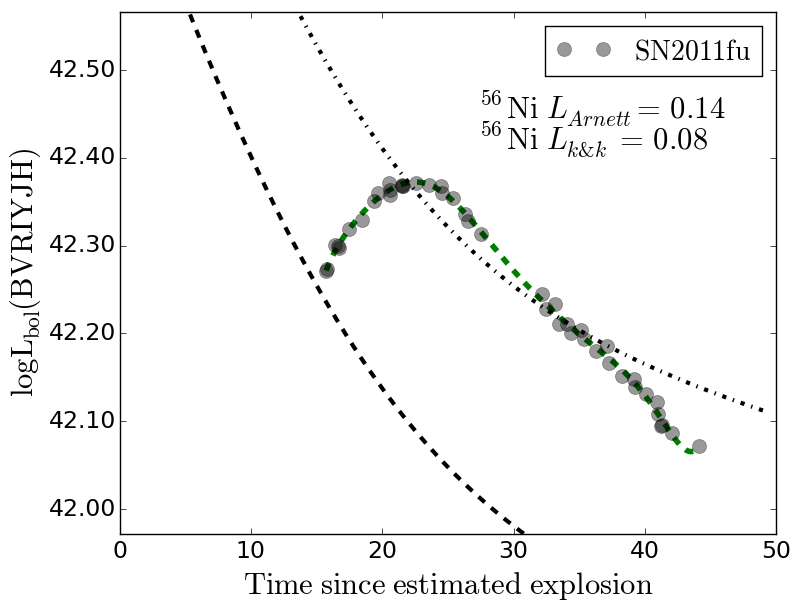}
\includegraphics[width=9cm]{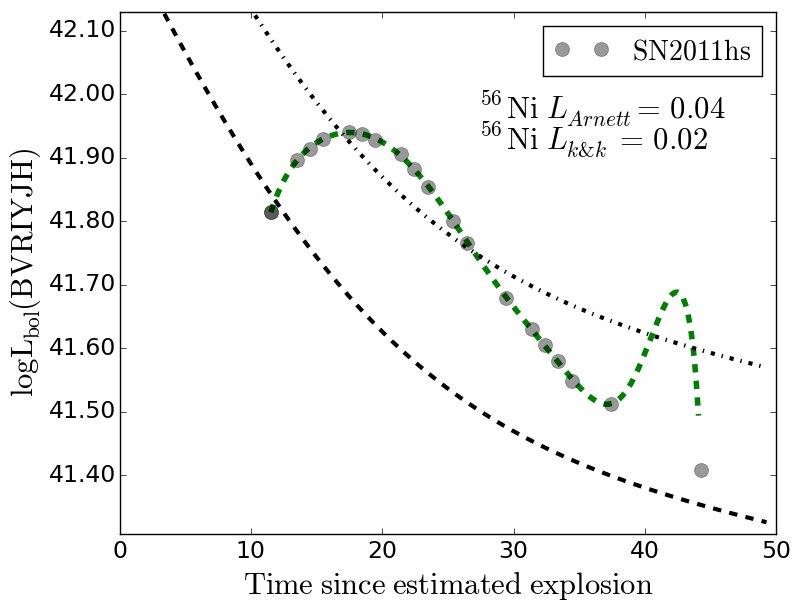}
\includegraphics[width=9cm]{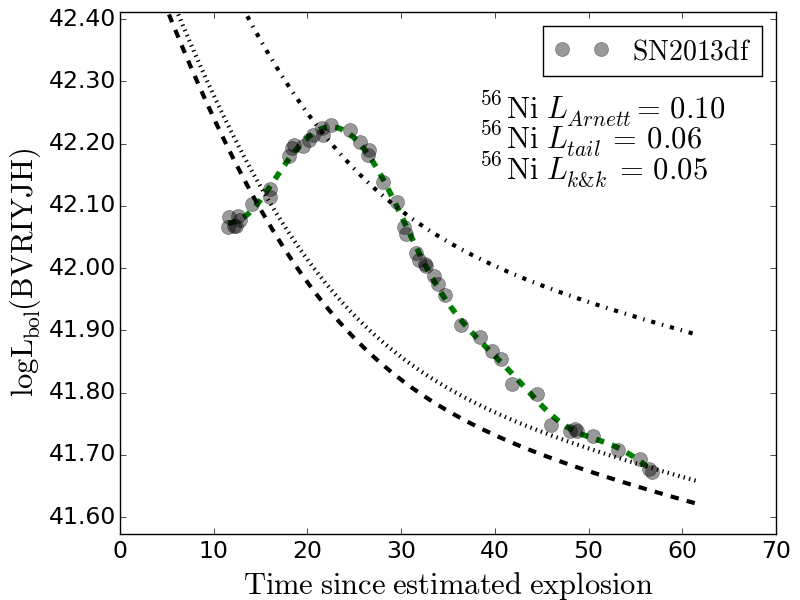}
\includegraphics[width=9cm]{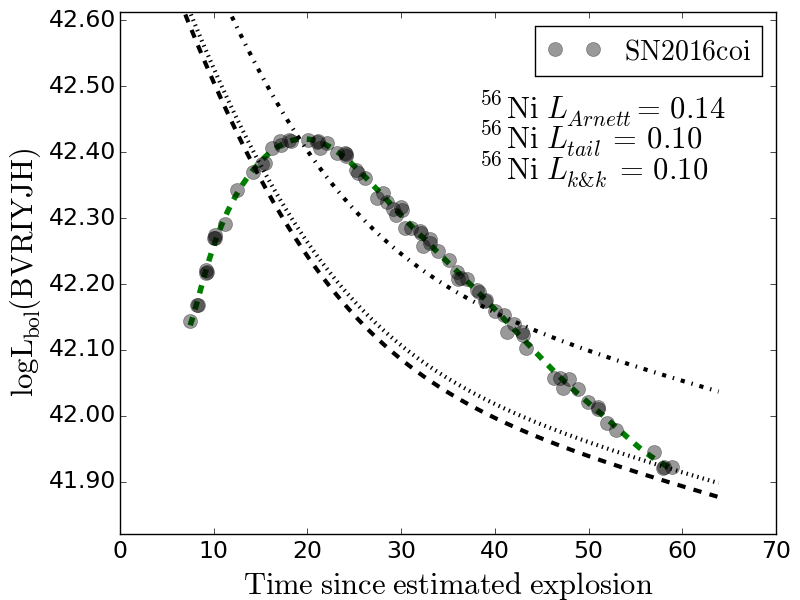}
\caption{$BVRIYJH$ pseudo-bolometric light curves for SE-SNe. The dotted lines give the $^{56}$Ni mass decay curve for that estimated through the Tail. The dashed line gives the $^{56}$Ni mass decay curve from \citeauthor{Khatami19}, while the dot-dashed line gives that from Arnett.}
\end{figure*}

\begin{figure}[ht!]
 \centering
\includegraphics[width=9cm]{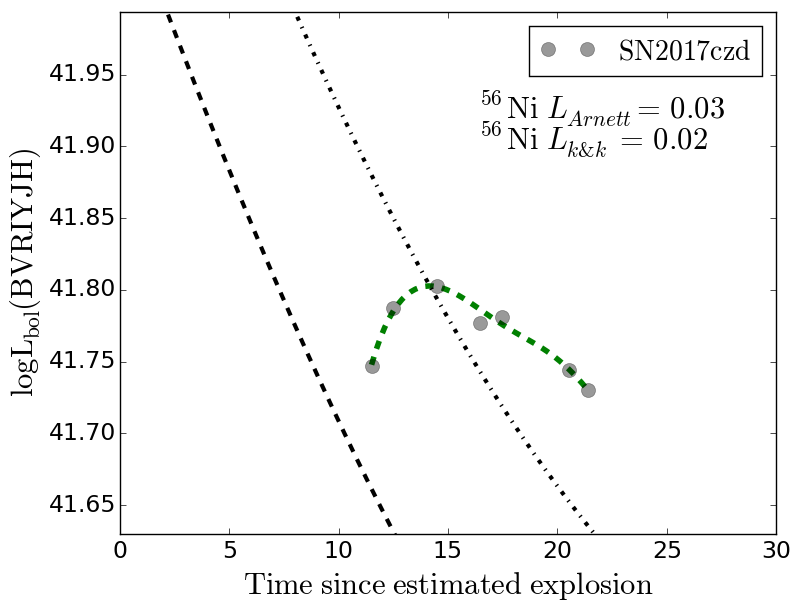}
\caption{$BVRIYJH$ pseudo-bolometric light curves for SE-SNe. The dotted lines give the $^{56}$Ni mass decay curve for that estimated through the Tail. The dashed line gives the $^{56}$Ni mass decay curve from \citeauthor{Khatami19}, while the dot-dashed line gives that from Arnett.}
\end{figure}

\end{document}